**The INSIDE assumption under all positive coding: interpretation and partial empirical assessment**

Fernando Pires Hartwig[1,2]*, George Davey Smith[2,3], Frank Dudbridge[4], Jack Bowden[3,5]

[1]Postgraduate Program in Epidemiology, Federal University of Pelotas, Pelotas, Brazil.

[2]MRC Integrative Epidemiology Unit, University of Bristol, Bristol, United Kingdom.

[3]Population Health Sciences, University of Bristol, Bristol, United Kingdom.

[4]Division of Public Health and Epidemiology, School of Medical Sciences, University of Leicester, Leicester, United Kingdom

[5]Department of Clinical Biosciences, Exeter Medical School, University of Exeter, Exeter, United Kingdom.

*Corresponding author. Postgraduate Program in Epidemiology, Federal University of Pelotas, Pelotas (Brazil) 96020-220. Phone: 55 53 32841300. E-mail: fernandophartwig@gmail.com.

## Abstract

Mendelian randomisation (MR) implemented through instrumental variable (IV) analysis is a popular strategy for strengthening causal inference in observational studies. A key assumption for several MR estimators, including MR-Egger regression, is the INstrument Strength Independent of Direct Effect (INSIDE) assumption. However, there is no established empirical test for assessing the plausibility of this assumption. Moreover, INSIDE depends on how genetic variants are coded (i.e., on the choice of the effect allele), which is often arbitrary and therefore hampers assessing the plausibility of this assumption on substantive grounds. In this paper, we show that the all-positive coding scheme (i.e., for all variants, choosing the allele positively associated with the exposure as the effect allele), which is typically used in MR-Egger, is equivalent to a coding-invariant model that can be given a natural interpretation because the direct effect parameters under this coding scheme are in the same direction as the bias of individual-variant ratio estimators. Moreover, using both theoretical arguments and simulations, we show that, under commonly assumed data-generating models in the MR methodological literature, heteroscedasticity of instrument-outcome coefficients according to instrument-exposure coefficients is a feature of at least some types of INSIDE violation, indicating that heteroscedasticity tests could contribute to assessing the plausibility of the INSIDE assumption. We further highlight specific cases where the test would not work. We illustrate its application by re-analysing a real dataset assessing the causal effect of large particle high density lipoprotein cholesterol on age-related macular degeneration.

## 1. Introduction

Mendelian randomisation (MR) leverages the special properties of germline genetic variation to strengthen causal inference using non-experimental data.[1,2] MR can be implemented through instrumental variables (IV) analysis, using genetic IVs to estimate the effect of modifying an exposure on some outcome of interest.[2,3] Valid causal inference using MR implemented through conventional IV analysis requires the following assumptions[2,3]: i) relevance: all genetic instruments are associated with the exposure (e.g., a disease risk factor); ii) independence: there are no common causes of the genetic instruments and the outcome (e.g., a disease endpoint); iii) exclusion restriction: any effect of the genetic instruments on the outcome is fully mediated by the exposure. Estimating a well-defined causal parameter (such as the average causal effect) also requires iv) additional parametric assumptions (such as linearity and homogeneity).[2-4] Since assumptions ii-iv cannot be fully empirically verified, there is a substantial body of research on developing estimators that relax one or more of these assumptions in different ways.

One such estimator is MR-Egger regression, which allows consistent causal effect estimation (provided that all other assumptions above are satisfied) even if all instruments violate the exclusion restriction assumption, provided that the INstrument Strength Independent of Direct Effect (INSIDE) assumption holds.[5] Since MR-Egger regression has been proposed, some tests to detect violations of the INSIDE assumption have been published, but have been shown to either partially test INSIDE by imposing additional assumptions (such as a zero mean exclusion restriction violation) or else employ circular reasoning.[6,7] To date, there is no established statistical approach for empirically assessing the plausibility of the INSIDE assumption.

Another important issue that has recently been raised with respect to MR-Egger is its dependence on allele coding. Typically, MR uses diallelic single nucleotide polymorphisms (SNPs) as candidate genetic instruments which are coded as the number of copies of the allele designated (often rather arbitrarily) as the effect allele. Although the choice of effect allele (i.e., the coding scheme) is irrelevant for several common MR methods, it is very relevant for MR-Egger regression because INSIDE may hold for some coding schemes but not others.[8,9] Traditionally, MR-Egger regression relies on the all-positive coding scheme – i.e., the allele estimated to be associated with higher levels of the exposure is chosen as the effect allele. This implies MR-Egger regression relies on a specific version of the INSIDE assumption: the all-positive coding version. This dependence of MR-Egger regression on a rather arbitrary issue potentially represents a major limitation of the method since assessing the arbitrary aspect of the required assumptions makes it difficult to assess their plausibility on substantive grounds.

In this paper, we further analyse these two limitations of MR-Egger regression. First, we show that the all-positive coding version of the INSIDE assumption is equivalent to a version of the INSIDE assumption that is coding-invariant and is naturally related to other common MR methods. Second, we propose a strategy that also allows partial testing for violation of this (or any other) version of the INSIDE assumption. We show that, under commonly assumed data-generating models in the two-sample MR literature, INSIDE violation will typically lead to heteroscedasticity. We also highlight special cases where this approach does not work and describe how the method can be further exploited to mitigate bias under additional assumptions. We then perform a series of simulations and illustrate the application of the method using a real data example assessing the causal effect of large particle high density lipoprotein cholesterol (XL-HDL-C) on age-related macular degeneration (AMD).

## 2. Data-generating model

We are interested in using $L$ mutually independent genetic variants (indexed by $j \in \mathbb{J} = \{1,2,\dots,L\}$) as IVs to estimate the causal effect of exposure $X$ on outcome $Y$. More specifically, we assume the following summary data-generating model:

$$\Gamma_j = \alpha_j + \gamma_j \beta \quad (1),$$

where $\gamma_j$ is the effect of $G_j$ (the $j$th genetic variant) on $X$, $\alpha_j$ is the direct effect of $G_j$ on $Y$ (i.e., the effect not mediated by $X$) and $\beta$ is the causal effect of $X$ on $Y$. For continuous $X$ and $Y$, $\gamma_j$ and $\Gamma_j$ typically correspond to per-effect allele average change in $X$ and $Y$, respectively. Different individual-level assumptions are sufficient for the model in equation (1) to be valid.[9-11] For example, if the effect of $X$ on $Y$ is additive linear (but not necessarily constant for all individuals in the studied population) and, for all $j \in \mathbb{J}$, the additive scale effects of $G_j$ on $X$ and of $X$ on $Y$ are uncorrelated (i.e., for all candidate genetic IVs, individuals for which the $G_j \rightarrow X$ effect is stronger do not have, on average, stronger or weaker $X \rightarrow Y$ effects), then the model in equation (1) holds and $\beta$ can be interpreted as the average causal effect.[11] Although this model makes several simplifications and may not be realistic in many empirical settings, it is nevertheless typically assumed in the methodological summary data MR literature, including MR-Egger regression.[5,9] Therefore, assuming this model does not impose additional assumptions beyond what are commonly assumed in the relevant literature.

Under the assumed summary data-generating model, the causal effect $\beta$ can be consistently estimated using $G_j$ if the following three assumptions hold:

- Relevance: $G_j$ and $X$ are correlated.
- Independence: $G_j$ and $Y$ have no common causes.
- Exclusion restriction: $G_j$ and $Y$ are independent conditional on $X$ and $U$ (where $U$ is the set of common causes of $X$ and $Y$). In the MR context, exclusion restriction can be violated due to horizontal pleiotropy (informally, “off-target” genetic effects).

Throughout this paper we assume that the relevance assumption holds – i.e., $\gamma_j \neq 0$ for all $j \in \mathbb{J}$. Violations of independence or exclusion restriction are modelled through the $\alpha_j$ term, so that $\alpha_j \neq 0$ indicates the $j$th variant is an invalid instrument.

The finite-sample estimates of $\gamma_j$ and $\Gamma_j$ are defined as follows:

$$\hat{\gamma}_j = \gamma_j + \varepsilon_{\gamma\,j}, \quad \mathrm{Var}\left(\varepsilon_{\gamma\,j}\right) = \sigma^2_{\gamma\,j} \quad (2).$$

$$\hat{\Gamma}_j = \Gamma_j + \varepsilon_{\Gamma\,j}, \quad \mathrm{Var}\left(\varepsilon_{\Gamma\,j}\right) = \sigma^2_{\Gamma\,j} \quad (3).$$

where $\sigma^2_{\gamma\,j}$ and $\sigma^2_{\Gamma\,j}$ are sampling variances that are assumed fixed and known. We assume $\hat{\gamma}_j$ and $\hat{\Gamma}_j$ are unbiased and consistent for $\gamma_j$ and $\Gamma_j$, respectively.

The parameter of interest (i.e., the populational quantity that one wants to consistently estimate) in this paper is $\beta$. We therefore refer to $\beta$ as the target parameter. We use the word estimand to refer to the asymptotic value of an estimator. For example, the ratio estimator corresponding to the $j$-th genetic variant is $\hat{\beta}_{R\,j} = \hat{\Gamma}_j/\hat{\gamma}_j$, and its asymptotic value is $\beta_{R\,j} = \Gamma_j/\gamma_j$ because $\hat{\gamma}_j$ and $\hat{\Gamma}_j$ are consistent. We therefore refer to $\beta_{R\,j}$ as the ratio estimand for the $j$-th variant.

## 3. MR-Egger and allele coding

### 3.1. Equivalence between the all-positive coding scheme and an allele-coding independent model

As shown elsewhere,[8,9] the INSIDE assumption $\mathrm{Cov}[\alpha_j, \gamma_j] = 0$ depends on how alleles are coded. This can lead to conceptual difficulties with respect to assessing the plausibility of the MR-Egger method, since this method uses a specific version of the INSIDE assumption – namely, the version under all positive coding (i.e., a coding scheme where, for all genetic variants, alleles are coded such that the exposure-increasing allele is considered the effect allele, thus $\gamma_j > 0$).

The recently proposed MR-GRIP model[9] achieves coding-invariance – the Genotype Recoding Invariance Property (GRIP) – by multiplying both sides of equation (1) by $\gamma_j$:

$$\Gamma_j \gamma_j = \alpha_j \gamma_j + \gamma_j^2 \beta \quad (4)$$

Notice $\gamma_j = \mathrm{sgn}(\gamma_j)|\gamma_j|$. Instead of $\gamma_j$, one can multiply both sides of equation (1) only by $\mathrm{sgn}(\gamma_j)$:

$$\mathrm{sgn}(\gamma_j)\Gamma_j = \mathrm{sgn}(\gamma_j)\alpha_j + |\gamma_j|\beta \quad (5),$$

which has GRIP. This is because, when swapping the effect and other alleles of $G_j$, the parameter corresponding to the effect of $G_j$ on $X$ $-\gamma_j$ and equation (1) becomes $-\Gamma_j =$ $-\alpha_j - \gamma_j\beta$, and multiplying both sides by $\text{sgn}(-\gamma_j)$ yields $\text{sgn}(-\gamma_j)(-\Gamma_j) = \text{sgn}(-\gamma_j)(-\alpha_j) + |-\gamma_j|\beta$, which is equivalent to equation (5).

Notably, equation (5) describes the same model produced by the all-positive coding scheme. That is, even though the all-positive coding scheme used in MR-Egger is a rather arbitrary choice out of all possible coding schemes, it is equivalent to a model that has the GRIP property. This does not mean that MR-Egger under all-positive coding intrinsically has GRIP, but rather that the all-positive coding scheme is mathematically equivalent to a model that has GRIP.

### 3.2. A coding-invariant interpretation of the version of the INSIDE assumption required by MR-Egger

This section uses the observation that the all-positive coding scheme is equivalent to a coding-invariant model to provide an interpretation of the INSIDE assumption under all-positive coding (the version of INSIDE that is required by MR-Egger) that is also coding invariant.

From equation (5), MR-Egger's causal effect estimand is $\beta_{Egger} = \text{Cov}[\text{sgn}(\gamma_j)\Gamma_j, |\gamma_j|]/\text{Var}[|\gamma_j|]$. If $\text{Cov}[\text{sgn}(\gamma_j)\alpha_j, |\gamma_j|] = 0$, then:

$$\begin{aligned}\beta_{Egger} &= \frac{\text{Cov}[\text{sgn}(\gamma_j)\Gamma_j, |\gamma_j|]}{\text{Var}[|\gamma_j|]} \\ &= \frac{\text{Cov}[\text{sgn}(\gamma_j)\alpha_j + |\gamma_j|\beta, |\gamma_j|]}{\text{Var}[|\gamma_j|]} \\ &= \frac{\text{Cov}[\text{sgn}(\gamma_j)\alpha_j, |\gamma_j|] + \text{Cov}[|\gamma_j|\beta, |\gamma_j|]}{\text{Var}[|\gamma_j|]} \\ &= \frac{0 + \beta\text{Cov}[|\gamma_j|, |\gamma_j|]}{\text{Var}[|\gamma_j|]} = \beta \quad (6)\end{aligned}$$

Therefore, $\text{Cov}[\text{sgn}(\gamma_j)\alpha_j, |\gamma_j|] = 0$, which is coding-invariant, is equivalent to the version of the INSIDE assumption required by MR-Egger – i.e., to the INSIDE assumption under all-positive coding.

For an intuitive interpretation of this coding-invariant version of the INSIDE assumption, consider $\beta_{R_j}$, which is obtained by dividing both sides of equation (1) by $\gamma_j$:

$$\beta_{R_j} = \frac{\Gamma_j}{\gamma_j} = \frac{\alpha_j + \gamma_j\beta}{\gamma_j} = b_j + \beta \quad (7),$$

where $b_j = \alpha_j/\gamma_j$ is the bias of $\beta_{R_j}$. It is well-known that $\beta_{R_j}$ has the GRIP property due to the division by $\gamma_j$.

By expressing the bias term as $b_j = \alpha_j/\gamma_j = \alpha_j \mathrm{sgn}(\gamma_j)/|\gamma_j| = \mathrm{sgn}(\alpha_j)\mathrm{sgn}(\gamma_j)|\alpha_j|/|\gamma_j|$, it is clear that the direct effect term in equation (6) has an immediate correspondence to the bias of the ratio estimand. Indeed, $\mathrm{sgn}(b_j) = \mathrm{sgn}(\alpha_j \mathrm{sgn}(\gamma_j)/|\gamma_j|) = \mathrm{sgn}\left(\alpha_j \mathrm{sgn}(\gamma_j)\right)$ – i.e., the direction of the direct effect term of the $j$-th genetic variant under all-positive coding is the same as the direction of the bias of the corresponding ratio estimand, whose magnitude $|b_j| = |\alpha_j/\gamma_j|$ is necessarily coding-invariant. Therefore, the direct effect terms under all-positive coding (which is equivalent to the coding-invariant model in equation (5)) and the bias terms of their corresponding ratio estimates have the same sign. Hereafter, we use the expression "bias-oriented direct effects" to denote $\alpha_j \mathrm{sgn}(\gamma_j)$ in equation (5).

The above implies a rather intuitive and natural interpretation of the direct effect terms:

- $\mathrm{Cov}(\alpha_j \mathrm{sgn}(\gamma_j), |\gamma_j|) > 0$: stronger instruments (i.e., those with larger $|\gamma_j|$) have on average more positively upwardly/less downwardly biasing direct effects than weaker instruments.

- $\mathrm{Cov}(\alpha_j \mathrm{sgn}(\gamma_j), |\gamma_j|) < 0$: stronger instruments have on average more downwardly/less upwardly biasing direct effects than weaker instruments.

- $\mathrm{Cov}(\alpha_j \mathrm{sgn}(\gamma_j), |\gamma_j|) = 0$: no correlation between instrument strength and bias-oriented direct effects.

## 4. Empirically assessing the plausibility of the INSIDE assumption

In this section, we discuss empirical strategies to assess the plausibility of the INSIDE assumption under all positive coding. To simplify the notation, let $\alpha_j^* = \alpha_j \mathrm{sgn}(\gamma_j)$, $\Gamma_j^* = \Gamma_j \mathrm{sgn}(\gamma_j)$ and $\gamma_j^* = |\gamma_j|$. Therefore, equation (5) can be written as:

$$\Gamma_j^* = \alpha_j^* + \gamma_j^* \beta \quad (8),$$

and the corresponding version of the INSIDE assumption (i.e., under all-positive coding) is $\mathrm{Cov}(\alpha_j^*, \gamma_j^*) = 0$.

### 4.1. The strong INSIDE assumption implies homoscedasticity

Consider a stronger version of the INSIDE assumption in which the parameters are independent rather than merely uncorrelated: $\alpha_j^* \perp\!\!\!\perp \gamma_j^*$, which we refer to as strong INSIDE. Under this assumption:

$$\begin{aligned}
\mathrm{Var}[\Gamma_j^* | \gamma_j^*] &= \mathrm{Var}[\alpha_j^* + \gamma_j^* \beta | \gamma_j^*] \\
&= \mathrm{Var}[\alpha_j^* | \gamma_j^*] + \mathrm{Var}[\gamma_j^* \beta | \gamma_j^*] + 2\mathrm{Cov}[\alpha_j^*, \gamma_j^* \beta | \gamma_j^*] \\
&= \mathrm{Var}[\alpha_j^*] \quad (9)
\end{aligned}$$

Therefore, strong INSIDE is sufficient for $\Gamma_j^*$ to be homoscedastic with respect to $\gamma_j^*$. This implies that, if $\Gamma_j^*$ is heteroscedastic with respect to $\gamma_j^*$, then strong INSIDE is violated – i.e., $\alpha_j^*$ and $\gamma_j^*$ are not independent. We will therefore discuss using heteroscedasticity tests as a means to assess the INSIDE assumption.

While it is clear that heteroscedasticity would be sufficient to falsify strong INSIDE, it is still possible that $\mathrm{Cov}(\alpha_j^*, \gamma_j^*) = 0$. This implies that falsifying strong INSIDE is not sufficient to falsify the original INSIDE assumption. Moreover, it is possible that strong INSIDE is violated, but there is no heteroscedasticity (e.g., the independence affects only higher-order moments). This implies that assessing heteroscedasticity cannot confirm strong INSIDE, which would be sufficient to confirm original INSIDE. Therefore, heteroscedasticity is neither sufficient nor necessary for the original INSIDE assumption to hold, which is a valid argument against the potential utility of heteroscedasticity assessment as a means of assessing the plausibility of INSIDE.

A possible response to this objection is that, even though strong INSIDE is technically stronger than needed for MR-Egger to be a consistent estimator of the causal effect, any biological justification of INSIDE would typically revolve around arguing for $\alpha_j^*$ (the bias-oriented direct effect) and $\gamma_j^*$ (instrument strength) being influenced by distinct biological pathways. Arguments of this nature, if true, would typically support strong INSIDE rather than just INSIDE. Indeed, such arguments are often qualitative in nature, so it seems rather unlikely for any biological justification of the INSIDE assumption to be specific about covariance.

To further illustrate this point, section 4.2 explores the implications of two distinct biological phenomena (horizontal pleiotropy and confounding between genetic instruments and $Y$) for both original and strong INSIDE.

## 4.2. Mechanistic motivation for assessing the plausibility of the INSIDE assumptions via heteroscedasticity tests

### 4.2.1. Horizontal pleiotropy

We will use the path diagram in Figure 1a to provide a mechanistic motivation for the utility of heteroscedasticity assessment for assessing the plausibility of the INSIDE assumption. In this diagram, $U_j$ denotes the set of confounders of the $X$-$Y$ association potentially affected by the $j$th instrument. This notation allows different variants to affect different confounders. However, this does not preclude the possibility that more than one instrument affects the same confounder: a scenario where $\kappa_{X_j} = \kappa_{X_{j'}}$ and $\kappa_{Y_j} = \kappa_{Y_{j'}}$ indicates that $G_j$ and $G_{j'}$ either affect the same confounders or affect confounders with the same effects on $X$ and $Y$.

According to this model:

$$\gamma_j^* = \delta_j + \psi_j \kappa_{X_j} \quad (10)$$

$$\Gamma_j^* = \alpha_{0\,j} + \delta_j\beta + \psi_j\kappa_{X\,j}\beta + \psi_j\kappa_{Y\,j}$$

$$= \left(\alpha_{0\,j} + \psi_j\kappa_{Y\,j}\right) + \left(\delta_j + \psi_j\kappa_{X\,j}\right)\beta \quad (11)$$

By comparing equations (8) and (11), it is clear that $\alpha_j^* = \alpha_{0\,j} + \psi_j\kappa_{Y\,j}$. In words, equation (10) describes that the total effect of $G_j$ on $X$ is a combination of the direct effect of $G_j$ on $X$ (i.e., the path $G_j \rightarrow X$ which corresponds to $\delta_j$) and the indirect effect mediated by $U_j$ (i.e., the path $G_j \rightarrow U_j \rightarrow X$ which corresponds to $\psi_j\kappa_{X\,j}$); and equation (11) describes that the total effect of $G_j$ on $Y$ is a combination of the direct effect of $G_j$ on $Y$ (i.e., the path $G_j \rightarrow Y$ which corresponds to $\alpha_{0\,j}$) and the indirect effects mediated by $X$ (i.e., the path $G_j \rightarrow X \rightarrow Y$ which corresponds to $\delta_j\beta$), by $U_j$ (i.e., the path $G_j \rightarrow U_j \rightarrow Y$ which corresponds to $\psi_j\kappa_{Y\,j}$) and by $U_j$ and $X$ (i.e., the path $G_j \rightarrow U_j \rightarrow X \rightarrow Y$ which corresponds to $\psi_j\kappa_{X\,j}\beta$).

We assume that direct genetic effect parameters $\delta_j$, $\alpha_{0\,j}$ and $\psi_j$ are mutually independent, and jointly independent on $\kappa_{X\,j}$ and $\kappa_{Y\,j}$. These assumptions imply that the only source of dependence between $\gamma_j^*$ and $\alpha_j^*$ is that $U_j$ mediates (some of) the effect of $G_j$ on $X$ and (some of) the effect of $G_j$ on $Y$ that is not mediated by $X$. Therefore, this path diagram combined with such independence assumptions allows isolating, and thus assessing the statistical implications of, a clear biological mechanism that could potentially induce INSIDE violations.

We note that:

$$\mathrm{Cov}\left[\gamma_j^*, \alpha_j^*\right] = \mathrm{Cov}\left[\delta_j + \psi_j\kappa_{X\,j}, \alpha_{0\,j} + \psi_j\kappa_{Y\,j}\right]$$

$$= \mathrm{Cov}\left[\psi_j\kappa_{X\,j}, \psi_j\kappa_{Y\,j}\right]$$

$$= \mathrm{E}\left[\psi_j\kappa_{X\,j}\psi_j\kappa_{Y\,j}\right] - \mathrm{E}\left[\psi_j\kappa_{X\,j}\right]\mathrm{E}\left[\psi_j\kappa_{Y\,j}\right]$$

$$= \mathrm{E}\left[\psi_j^2\right]\mathrm{E}\left[\kappa_{X\,j}\kappa_{Y\,j}\right] - \left(\mathrm{E}\left[\psi_j\right]\right)^2\mathrm{E}\left[\kappa_{X\,j}\right]\mathrm{E}\left[\kappa_{Y\,j}\right]$$

$$= \mathrm{E}\left[\psi_j^2\right]\mathrm{Cov}\left[\kappa_{X\,j}, \kappa_{Y\,j}\right] + \mathrm{Var}\left[\psi_j\right]\mathrm{E}\left[\kappa_{X\,j}\right]\mathrm{E}\left[\kappa_{Y\,j}\right] \quad (12)$$

Therefore, without making assumptions on $\kappa_{X\,j}$ and $\kappa_{Y\,j}$ (more specifically, on their means and covariance), $\mathrm{Cov}\left[\gamma_j^*, \alpha_j^*\right] = 0$ in general only if $\psi_j = 0$ for all $j \in \mathbb{J}$. Moreover:

$$\mathrm{Var}\left[\Gamma_j^*|\gamma_j^*\right] = \mathrm{Var}\left[\alpha_j^*|\gamma_j^*\right]$$

$$= \mathrm{Var}\left[\alpha_{0\,j} + \psi_j\kappa_{Y\,j}|\gamma_j^*\right]$$

$$= \mathrm{Var}\left[\alpha_{0\,j}\right] + \mathrm{Var}\left[\psi_j\kappa_{Y\,j}|\gamma_j^*\right] \quad (13)$$

Since $\psi_j$ is in the right-hand side of $\gamma_j^*$, equation (13) shows that, without further assumptions, in general $\text{Var}[\Gamma_j^*|\gamma_j^*]$ will be a function of $\gamma_j^*$ – i.e., $\Gamma_j^*$ will be heteroscedastic with respect to $\gamma_j^*$. Importantly, if $\psi_j = 0$ for all $j \in \mathbb{J}$, then $\text{Var}[\Gamma_j^*|\gamma_j^*] = \text{Var}\left[\alpha_{0_j}\right]$ – i.e., $\Gamma_j$ is homoscedastic with respect to $\gamma_j^*$. Therefore, heteroscedasticity would imply $\psi_j \neq 0$ for at least one $j \in \mathbb{J}$, which in turn implies $\text{Cov}[\gamma_j^*, \alpha_j^*] \neq 0$ in general, unless one is willing to make assumptions about the direct effects of unmeasured and potentially unknown variables affecting both $X$ and $Y$. Indeed, it is in circumstances where such knowledge is unavailable that causal effect estimation using instrumental variables is of greater interest.

Note that $\psi_j \neq 0$ for at least one $j \in \mathbb{J}$ does not necessarily imply heteroscedasticity even under the assumptions maintained in this section. For example, suppose $\psi_j = c \neq 0$ for all $j \in \mathbb{J}$. Then, $\text{Cov}[\gamma_j^*, \alpha_j^*] \neq 0$ if $\text{Cov}\left[\kappa_{X_j}, \kappa_{Y_j}\right] \neq 0$ (by equation (12)) but $\text{Var}\left[\psi_j \kappa_{Y_j} | \gamma_j^*\right]$ does not depend on $\gamma_j^*$, thus implying $\Gamma_j^*$ is homoscedastic with respect to $\gamma_j^*$ (by equation (13)). This illustrates that, in the model considered in this section, heteroscedasticity requires $\text{Var}[\psi_j] \neq 0$, which implies $\psi_j \neq 0$ for at least one $j \in \mathbb{J}$.

### 4.2.2. Confounding between genetic variants and the outcome

Figure 1a showed a path diagram where genetic variants had direct effects on the outcome. We now consider a different mechanism: unmeasured confounding between the genetic variants and the outcome, as illustrated in Figure 1b. In this diagram, $W_j$ denotes the set of unmeasured common causes of $G_j$ and $Y$ (which may or may not also affect $X$). As in section 4.2.1, this notation allows different variants to be affected by different confounders without precluding the possibility that more than one instrument is affected by the same confounder.

According to this model:

$$\gamma_j^* = \delta_j + \pi_{G_j} \pi_{X_j} \text{Var}[W_j]/\text{Var}[G_j] \quad (14)$$

$$\Gamma_j^* = \pi_{G_j} \pi_{Y_j} \text{Var}[W_j]/\text{Var}[G_j] + \left(\delta_j + \pi_{G_j} \pi_{X_j} \text{Var}[W_j]/\text{Var}[G_j]\right) \beta \quad (15)$$

By comparing equations (8) and (15), it is clear that $\alpha_j^* = \pi_{G_j} \pi_{Y_j} \text{Var}[W_j]/\text{Var}[G_j]$. In words, equation (14) describes that the $G_j$-$X$ coefficient is a combination of the effect of $G_j$ on $X$ (i.e., the path $G_j \to X$ which corresponds to $\delta_j$) and $W_j$-induced confounding (i.e., the path $G_j \leftarrow W_j \to X$ which corresponds to $\pi_{G_j} \pi_{X_j} \text{Var}[W_j]/\text{Var}[G_j]$); and equation (15) describes that the $G_j$-$Y$ coefficient is a combination of the effect of $G_j$ on $Y$ (i.e., the path $G_j \to X \to Y$ which corresponds to $\delta_j \beta$), $W_j$-induced confounding mediated by $X$ (i.e., the $G_j \leftarrow W_j \to X \to Y$) and direct $W_j$-induced confounding (i.e., the path $G_j \leftarrow W_j \to X \to Y$).

We assume that $\delta_j$ and $\mathrm{Var}\left[W_j\right]/\mathrm{Var}\left[G_j\right] \pi_{G\,j} = \tau_j$ (i.e., the parameters corresponding to the effects on/of $G_j$) are mutually independent, and jointly independent on $\pi_{X\,j}$ and $\pi_{Y\,j}$. These assumptions imply that the only source of dependence between $\gamma_j^*$ and $\alpha_j^*$ is that $W_j$-induced confounding influences both the $G_j$-$X$ and the $G_j$-$Y$ coefficients in a way that is not solely mediated by $X$. Therefore, this path diagram combined with such independence assumptions allows isolating, and thus assessing the statistical implications of, a second clear biological mechanism that could potentially induce INSIDE violations.

We note that:

$$
\begin{aligned}
\mathrm{Cov}\left[\gamma_j^*, \alpha_j^*\right] &= \mathrm{Cov}\left[\delta_j + \pi_{G\,j}\pi_{X\,j}\,\mathrm{Var}\left[W_j\right]/\mathrm{Var}\left[G_j\right], \pi_{G\,j}\pi_{Y\,j}\,\mathrm{Var}\left[W_j\right]/\mathrm{Var}\left[G_j\right]\right] \\
&= \mathrm{Cov}\left[\delta_j + \tau_j\pi_{X\,j}, \tau_j\pi_{Y\,j}\right] \\
&= \mathrm{Cov}\left[\tau_j\pi_{X\,j}, \tau_j\pi_{Y\,j}\right] \\
&= \mathrm{E}\left[\tau_j^2\pi_{X\,j}\pi_{Y\,j}\right] - \mathrm{E}\left[\tau_j\pi_{X\,j}\right]\mathrm{E}\left[\tau_j\pi_{Y\,j}\right] \\
&= \mathrm{E}\left[\tau_j^2\right]\mathrm{E}\left[\pi_{X\,j}\pi_{Y\,j}\right] - \left(\mathrm{E}\left[\tau_j\right]\right)^2\mathrm{E}\left[\pi_{X\,j}\right]\mathrm{E}\left[\pi_{Y\,j}\right] \\
&= \mathrm{E}\left[\tau_j^2\right]\mathrm{Cov}\left[\pi_{X\,j}\pi_{Y\,j}\right] + \mathrm{Var}\left[\tau_j\right]\mathrm{E}\left[\pi_{X\,j}\right]\mathrm{E}\left[\pi_{Y\,j}\right] \qquad (16)
\end{aligned}
$$

Therefore, without making assumptions on $\pi_{X\,j}$ and $\pi_{Y\,j}$ (more specifically, on their means and covariance), $\mathrm{Cov}\left[\gamma_j^*, \alpha_j^*\right] = 0$ in general only if $\tau_j = 0$ (which implies either $\pi_{G\,j} = 0$ or $\mathrm{Var}\left[W_j\right] = 0$) for all $j \in \mathbb{J}$, in which case there is no unmeasured $G_j$-$Y$ confounding that also affects $X$. Indeed, if $\pi_{X\,j} = 0$ for all $j \in \mathbb{J}$, then $\mathrm{Cov}\left[\pi_{X\,j}, \pi_{Y\,j}\right] = \mathrm{E}\left[\pi_{X\,j}\right] = 0$, implying $\mathrm{Cov}\left[\gamma_j^*, \alpha_j^*\right] = 0$. That is, unmeasured $G_j$-$Y$ confounding that does not affect $X$ does not lead to INSIDE violations. Moreover:

$$
\begin{aligned}
\mathrm{Var}\left[\Gamma_j^* | \gamma_j^*\right] &= \mathrm{Var}\left[\alpha_j^* | \gamma_j^*\right] \\
&= \mathrm{Var}\left[\tau_j\pi_{Y\,j} | \gamma_j^*\right] \qquad (17)
\end{aligned}
$$

Since $\gamma_j^* = \delta_j + \tau_j\pi_{X\,j}$, $\tau_j$ is in the right-hand side of both $\alpha_j^*$ and $\gamma_j^*$. Therefore, without further assumptions, $\Gamma_j^*$ will in general be heteroscedastic with respect to $\gamma_j$. Importantly, if $\tau_j = 0$ for all $j \in \mathbb{J}$, then $\mathrm{Var}\left[\Gamma_j^* | \gamma_j^*\right] = 0$ – i.e., $\Gamma_j$ is constant (and therefore homoscedastic) with respect to $\gamma_j^*$. Therefore, heteroscedasticity would imply $\tau_j \neq 0$ for at least one $j \in \mathbb{J}$, which in turn implies $\mathrm{Cov}\left[\gamma_j^*, \alpha_j^*\right] \neq 0$ in general, unless one is willing to make assumptions about the direct effects of unmeasured and potentially unknown variables affecting $G_j$, $X$ and $Y$. As mentioned above, it is in circumstances where such knowledge is unavailable that causal effect estimation using instrumental variables is of greater interest. Of note, if $\pi_{X_j} = 0$ for all $j \in \mathbb{J}$, then $\text{Var}\left[\tau_j \pi_{Y_j} | \gamma_j^*\right] = \text{Var}\left[\tau_j \pi_{Y_j} | \delta_j\right] = \text{Var}\left[\tau_j \pi_{Y_j}\right]$, showing that unmeasured $G_j$-$Y$ confounding that does not affect $X$ does not lead to heteroscedasticity.

Note that $\tau_j \neq 0$ for at least one $j \in \mathbb{J}$ does not necessarily imply heteroscedasticity even under the assumptions maintained in this section. For example, suppose $\tau_j = c \neq 0$ for all $j \in \mathbb{J}$. Then, $\text{Cov}\left[\gamma_j^*, \alpha_j^*\right] \neq 0$ if $\text{Cov}\left[\kappa_{X_j}, \kappa_{Y_j}\right] \neq 0$ (by equation (16)) and $\text{Var}\left[\tau_j \pi_{Y_j} | \gamma_j^*\right]$ does not depend on $\gamma_j^*$, thus implying $\Gamma_j^*$ is homoscedastic with respect to $\gamma_j^*$ (by equation (17)). This illustrates that, in the model considered in this section, heteroscedasticity requires $\text{Var}\left[\tau_j\right] \neq 0$, which implies $\tau_j \neq 0$ for at least one $j \in \mathbb{J}$.

### 4.2.3. Reverse causation

In this section we consider yet another mechanism: reverse causation (i.e., $Y$ causes $X$). This is illustrated in the path diagram in Figure 1c by the arrow from $Y$ to $X$ (notice that in this diagram we use $\beta'$ instead of $\beta$ to represent the effect of $Y$ on $X$ rather than the effect of $X$ on $Y$). In this diagram, $W_j$ denotes the set of unmeasured common causes of $G_j$ and $Y$ (which may or may not also affect $X$). Since no candidate genetic instrument causes or is caused by $X$-$Y$ confounders, there is a single $U$ denoting the set of such confounders.

According to this model:

$$\gamma_j^* = \delta_j + \alpha_{0_j} \beta' \quad (18)$$

$$\Gamma_j^* = \alpha_{0_j} \quad (19)$$

By comparing equations (8) and (19), it is clear that $\alpha_j^* = \alpha_{0_j}$. In words, equation (18) describes that the total effect of $G_j$ on $X$ is a combination of the direct effect of $G_j$ on $X$ (i.e., the path $G_j \rightarrow X$ which corresponds to $\delta_j$) and the indirect effect mediated by $Y$ (i.e., the path $G_j \rightarrow Y \rightarrow X$ which corresponds to $\alpha_{0_j} \beta'$); and equation (19) describes that the total effect of $G_j$ on $Y$ is simply the direct effect of $G_j$ on $Y$ (i.e., the path $G_j \rightarrow Y$ which corresponds to $\alpha_{0_j}$).

We assume that direct genetic effect parameters $\delta_j$ and $\alpha_{0_j}$ are independent. This assumption implies that the only source of dependence between $\gamma_j^*$ and $\alpha_j^*$ is that both include $\alpha_{0_j}$, which is a consequence of $Y$ being a cause of of $X$. Therefore, this path diagram combined with this independence assumption allows isolating, and thus assessing the statistical implications of, a third clear biological mechanism that could potentially induce INSIDE violations.

We note that:

$$\mathrm{Cov}[\gamma_j^*, \alpha_j^*] = \mathrm{Cov}\left[\delta_j + {\alpha_0}_j\beta', {\alpha_0}_j\right]$$

$$= \beta'\mathrm{Var}\left[{\alpha_0}_j\right] \quad (20)$$

Therefore, if there is reverse causation (i.e., $\beta' \neq 0$), then $\mathrm{Cov}[\gamma_j^*, \alpha_j^*] = 0$ if and only if $\mathrm{Var}\left[{\alpha_0}_j\right] = 0$, which is trivial in this case because ${\alpha_0}_j = \alpha_j^*$, so $\mathrm{Var}\left[{\alpha_0}_j\right] = 0$ implies $\mathrm{Var}[\alpha_j^*] = 0$. Moreover:

$$\mathrm{Var}[\Gamma_j^*|\gamma_j^*] = \mathrm{Var}[\alpha_j^*|\gamma_j^*]$$

$$= \mathrm{Var}\left[{\alpha_0}_j|\gamma_j^*\right] \quad (21)$$

Since ${\alpha_0}_j$ is in the right-hand side of $\gamma_j^*$, equation (21) shows that, without further assumptions, in general $\mathrm{Var}[\Gamma_j^*|\gamma_j^*]$ will be a function of $\gamma_j^*$ – i.e., $\Gamma_j^*$ will be heteroscedastic with respect to $\gamma_j^*$. Importantly, if $\mathrm{Var}\left[{\alpha_0}_j\right] = 0$, then $\mathrm{Var}[\Gamma_j^*|\gamma_j^*] = 0$ – i.e., $\Gamma_j$ is homoscedastic with respect to $\gamma_j^*$. Therefore, heteroscedasticity would imply $\mathrm{Var}\left[{\alpha_0}_j\right] > 0$, which in turn implies $\mathrm{Cov}[\gamma_j^*, \alpha_j^*] \neq 0$ unless $\beta' = 0$.

### 4.3. Assessing the plausibility of the VICE assumption via heteroscedasticity assessment

From equation (4), MR-GRIP's causal effect estimand is $\beta_{GRIP} = \mathrm{Cov}[\Gamma_j\gamma_j, \gamma_j^2]/\mathrm{Var}[\gamma_j^2]$. If $\mathrm{Cov}[\gamma_j\alpha_j, \gamma_j^2] = 0$, then:

$$\beta_{GRIP} = \frac{\mathrm{Cov}[\Gamma_j\gamma_j, \gamma_j^2]}{\mathrm{Var}[\gamma_j^2]}$$

$$= \frac{\mathrm{Cov}[\gamma_j\alpha_j + \gamma_j^2\beta, \gamma_j^2]}{\mathrm{Var}[\gamma_j^2]}$$

$$= \frac{\mathrm{Cov}[\gamma_j\alpha_j, \gamma_j^2] + \mathrm{Cov}[\gamma_j^2\beta, \gamma_j^2]}{\mathrm{Var}[\gamma_j^2]}$$

$$= \frac{0 + \beta\mathrm{Cov}[\gamma_j^2, \gamma_j^2]}{\mathrm{Var}[\gamma_j^2]} = \beta \quad (22)$$

Therefore, $\mathrm{Cov}[\gamma_j\alpha_j, \gamma_j^2] = 0$, which is coding-invariant, is sufficient for $\beta_{GRIP} = \beta$. This has been referred to as the VICE assumption. A stronger version of this assumption (which we refer to as strong VICE) would be $\alpha_j\gamma_j \perp\!\!\!\perp \gamma_j^2$ (see section 4.1 for some justification for considering a stronger version this assumption).

Under strong VICE:

$$\mathrm{Var}\left[\Gamma_j\gamma_j|\gamma_j^2\right] = \mathrm{Var}\left[\alpha_j\gamma_j + \gamma_j^2\beta|\gamma_j^2\right]$$

$$= \mathrm{Var}\left[\alpha_j\gamma_j|\gamma_j^2\right] + \mathrm{Var}\left[\gamma_j^2\beta|\gamma_j^2\right] + 2\mathrm{Cov}\left[\alpha_j\gamma_j, \gamma_j^2\beta|\gamma_j^2\right]$$

$$= \mathrm{Var}\left[\alpha_j\gamma_j\right] \quad (23)$$

Therefore, strong VICE is sufficient for $\Gamma_j\gamma_j$ to be homoscedastic with respect to $\gamma_j^2$. This implies that, if $\Gamma_j\gamma_j$ is heteroscedastic with respect to $\gamma_j^2$, then strong VICE is violated – i.e., $\alpha_j\gamma_j$ and $\gamma_j^2$ are not independent, which would at least question the validity of VICE.

### 4.4. Limitations of testing for heteroscedasticity as an INSIDE test

The above suggests that, if there is evidence against the null hypothesis that homoscedasticity holds – i.e., if $H_0\colon \mathrm{Var}\left[\Gamma_j^*|\gamma_j^*\right] = \mathrm{Var}\left[\Gamma_{j'}^*|\gamma_{j'}^*\right]$ for all for all $j, j' \in \mathbb{J}$ is unlikely – then maintaining the INSIDE assumption would be unwarranted. However, we also mentioned that this test would fail in some circumstances, which we further discuss below.

#### 4.4.1. Situations where $H_0$ is false, but INSIDE holds

Testing the above $H_0$ can, in some circumstances, lead to a false indication that there is bias due to INSIDE violation. This is simply because $H_0$ being false does not imply $\mathrm{Cov}\left[\gamma_j^*, \alpha_j^*\right] = 0$. Therefore, it is possible that $H_0$ is false and $\mathrm{Cov}\left[\gamma_j^*, \alpha_j^*\right] = 0$, implying the causal effect estimator would be consistent for $\beta$. In this situation, the fact that $H_0$ is false would be classified as a false-positive result. Given the model assumed in section 4.2.1, if $H_0$ is false, then at least one candidate genetic variant is an invalid IV. So, a situation where $H_0$ is false and $\mathrm{Cov}\left[\gamma_j^*, \alpha_j^*\right] = 0$ would be a situation of balanced horizontal pleiotropy.

Another potential issue concerns the fact that, in practice, only finite-sample estimates $\hat{\gamma}_j^*$ and $\hat{\Gamma}_j^*$ are available. Therefore, it is possible that $H_0$ is rejected because $\sigma_{\Gamma\ j}^2 \neq \sigma_{\Gamma\ j'}^2$ for some $j$, $j' \in \mathbb{J}$, where $\hat{\gamma}_j^* \neq \hat{\gamma}_{j'}^*$ (i.e., if $\sigma_{\Gamma\ j}^2$ is not constant along instrument strength values). To illustrate, consider a situation in which $\sigma_{\Gamma\ j}^2 > \sigma_{\Gamma\ j'}^2$ for $\hat{\gamma}_j^* > \hat{\gamma}_{j'}^*$ – i.e., the sampling variance of $\hat{\Gamma}_j^*$ is larger for stronger instruments. In this situation, $H_0$ would be rejected by applying statistical tests to finite-sample estimates even if $H_0$ holds.

One example where this could happen is if $Y$ is binary and $\hat{\Gamma}_j^*$ is expressed as ln(odds ratio). In this case, $\sigma_{\Gamma\ j}^2$ will depend on $\hat{\gamma}_j^*$ because the sampling error and the effect estimate will not be independent. Importantly, this condition can be empirically assessed by testing the correlation between $\hat{\gamma}_j^*$ and $\hat{\sigma}_{\Gamma\ j}^2$. This is illustrated in section 5.

#### 4.4.2. Situations where $H_0$ is true, but INSIDE is violated

As mentioned in section 4.2.1, testing $H_0$ would only detect INSIDE violation due to horizontal pleiotropy if $\text{Var}[\psi_j] > 0$. If $\psi_j = c \neq 0$ for all $j \in \mathbb{J}$, then $H_0$ would hold even though INSIDE is violated. Although this situation is possible, it is quite contrived. First, it requires that all variants have a direct effect on a common cause of $X$ and $Y$. Second, it requires that such violations are identical for all variants, which is especially unlikely when many genetic variants in different genes affecting multiple biological pathways are used as instruments.

Similarly, as mentioned in section 4.2.2, testing $H_0$ would only detect INSIDE violation due to unmeasured confounding between genetic variants and the outcome if $\text{Var}[\tau_j] > 0$. If $\tau_j = c \neq 0$ for all $j \in \mathbb{J}$, then $H_0$ would hold even though INSIDE is violated.

## 4.5. Bias correction

As discussed above, the plausibility of the INSIDE assumption is questionable in the presence of heteroscedasticity. Therefore, a natural question is whether this can be explored not only for (risk of) bias detection, but also for bias correction. Given the limitations discussed in section 4.4, this is not possible without further assumptions. A sufficient extra assumption for this purpose can be described as follows. First, let $S_k$ be one of the $K = 2^L - \sum_{r=0}^{2} \binom{L}{r}$ subsets of variants containing at least three variants (for less than three variants, heteroscedasticity cannot be detected after fitting a linear model as in MR-Egger regression). Let $Q_k \in \{0,1\}$ indicate whether the $k$-th subset is homoscedastic ($Q_k = 1$) or not ($Q_k = 0$). Finally, let $k^*$ denote the largest $S_k$ such that $Q_k = 1$, which we assume to be unique (i.e., there is a single largest homoscedastic subset of variants). We assume that INSIDE holds in the $k^*$-th subset – i.e., $\text{Cov}[\gamma_j^*, \alpha_j^* | j \in S_{k^*}] = 0$. A conceptually simple way to implement this strategy is to exhaustively search through all possible subsets of the $L$ genetic variants to identify the largest one where homoscedasticity holds. However, this is not computationally feasible when $L$ is relatively large. In section 5 we illustrate how this approach can be used in practice when removing a single variant strongly attenuates heteroscedasticity. We discuss the issue of the computational burden of the approach for more extensive searches over subsets of variants in more detail in section 6.

This assumption is somewhat analogous to the plurality rule (also referred to as the ZEro Modal Pleiotropy Assumption – ZEMPA),[12,13] which postulates that the largest subset of variants with common ratio estimands consists of variants satistfying the IV assumptions (in our notation, this can be expressed as $\text{mode}(b_j) = 0$; this assumption can be straightforwardly generalized to accommodate weighted versions). The assumption outlined above postulates that the largest subset of homoscedastic variants consists of variants satisfying INSIDE. Importantly, without biological justification for the choice of the genetic variants, the plurality rule is not necessarily plausible. In this case, its utility will typically be relaxing other identification assumptions, such as the conventional assumption that $b_j = 0$ for all $j \in \mathbb{J}$; and, at least in the case of $\text{sgn}(b_j) = \text{sgn}(b_{j'})$ for all $j, j' \in \mathbb{J}$, assumptions such as the majority rule ($\text{median}(b_j) = 0$). Similarly, without biological justification, the assumption $\text{Cov}[\gamma_j^*, \alpha_j^* | j \in S_{k^*}] = 0$ serves mainly to relax the original INSIDE assumption from a statistical perspective.

## 4.6. Heterogeneity versus heteroscedasticity

Testing for homogeneity between ratio estimates (that is, testing the null hypothesis that $\beta_{R_1} = \beta_{R_2} = \cdots = \beta_{R_L}$) has been proposed as a potential way to assess the plausibility of no horizontal pleiotropy in the context of two-sample MR.[10,14] This exploits the assumption that ratio estimands can only vary due to bias, as assumed here. Of course, this may not hold in practice, and thus the test can reject the null hypothesis of homogeneity even if relevance, independence and exclusion restriction all hold for all instruments. Nevertheless, with these limitations in mind, the test can be useful, in combination with other approaches, for empirically exploring the possibility of horizontal pleiotropy.

In this section we clarify the distinctions between heterogeneity and heteroscedasticity tests for the purposes of assessing horizontal pleiotropy in two-sample MR. However, it is important to note that the tests have several commonalities. First, both require the causal effect homogeneity assumption (or at least some sufficient assumption for all valid instruments to have the same ratio estimand) for rejection of the null hypothesis to be interpreted as evidence of horizontal pleiotropy. Second, neither of the tests quantify the bias. Third, it is possible for both tests that the null hypothesis is rejected but there is no bias (the case of "balanced" horizontal pleiotropy) and that the null hypothesis is not rejected but there is bias (the case of homogeneous biases across all variants).

Even though there are similarities, the tests have a fundamental difference, which we discuss below under the data-generating model of section 4.2.1:

- Heterogeneity does not imply heteroscedasticity: this can be shown by example. Assuming $\psi_j = 0$ and $\alpha_{0_j} = c \neq 0$ for all $j \in \mathbb{J}$, equation (14) implies $\text{Var}[\Gamma_j^* | \gamma_j^*] = 0$, which is clearly not a function of $\gamma_j^*$ – i.e., homoscedasticity holds. Moreover, $\beta_{R_j} = \frac{c}{\gamma_j} + \beta \neq \frac{c}{\gamma_{j'}} + \beta = \beta_{R_{j'}}$. Therefore, if $\text{Var}[\gamma_j^*] > 0$, then $\beta_{R_j} \neq \beta_{R_{j'}}$ for at least one pair $j, j' \in \mathbb{J}$ – i.e., homogeneity does not hold. Therefore, there is heterogeneity but not heteroscedasticity, showing that heterogeneity does not necessarily imply heteroscedasticity.

- Heteroscedasticity implies heterogeneity: from equation (14), heteroscedasticity requires $\text{Var}\left[\psi_j \kappa_{Y_j}\right] > 0$. From equation (8) and the independence assumptions described in section 4.2.1, $\gamma_j^* = \delta_j + \psi_j \kappa_{X_j}$ and $\alpha_j^* = \alpha_{0_j} + \psi_j \kappa_{Y_j}$ are independent, implying $\text{Var}[b_j] > 0$ (i.e., by independence, variability in $\psi_j \kappa_{Y_j}$ is not expected to be "cancelled out" by adding $\alpha_{0_j}$ and dividing the sum by $\gamma_j^*$). Since $b_j$ is the only source of variability between ratio estimands given the assumption of causal effect homogeneity, $\text{Var}[b_j] > 0$ implies $\text{Var}\left[\beta_{R_j}\right] > 0$. Therefore, heteroscedasticity implies heterogeneity in this data-generating model.

This difference between heterogeneity and heteroscedasticity makes intuitive sense: heterogeneity tests aim at detecting any form of non-constant horizontal pleiotropy; while heteroscedasticity tests aim at detecting only forms of pleiotropy that depend on $\gamma_j^*$. That is, in this application, heterogeneity tests are less specific than heteroscedasticity tests, detecting more forms of horizontal pleiotropy than heteroscedasticity tests do.

## 5. Simulation study

### 5.1. Data-generating model

We performed a simulation study to corroborate the theoretical results. The simulations were performed using the data-generating model described in section 4.2.1. Additional details of the model are described below.

Let $n_X$ and $n_Y$ denote the sample sizes used to estimate $\hat{\gamma}_j^* = |\hat{\gamma}_j|$ and $\hat{\Gamma}_j^* = \hat{\Gamma}_j \text{sgn}(\hat{\gamma}_j)$. $\theta_j \sim \text{Uniform}(0.1,0.9)$ denotes the effect allele frequency. We also assume that $\text{Var}[X] = \text{Var}[Y] = 1$. From this, assuming that ordinary least squares estimation is used, it follows that $\sigma_{\gamma\,j}^2 = \frac{1-2\theta_j(1-\theta_j)\gamma_j^2}{2\theta_j(1-\theta_j)n_X}$ and $\sigma_{\Gamma\,j}^2 = \frac{1-2\theta_j(1-\theta_j)\Gamma_j^2}{2\theta_j(1-\theta_j)n_Y}$. Finite-sample estimates were generated assuming independent Normal distributions: $\hat{\gamma}_j \sim N\left(\gamma_j, \sigma_{\gamma\,j}^2\right)$ and $\hat{\Gamma}_j \sim N\left(\Gamma_j, \sigma_{\Gamma\,j}^2\right)$. This corresponds to the two-sample MR setting where $\hat{\gamma}_j$ and $\hat{\Gamma}_j$ are estimated in non-overlapping samples and are therefore independent.

Defining the model in this way helps to clarify its interpretation. For example, $\gamma_j$ can be interpreted as the average change in $X$ in standard deviation units per-allele increment in $G_j$, and $\beta$ can be interpreted as the causal Pearson correlation coefficient (an interpretation that also holds for $k_{X\,j} = k_{Y\,j} \sim \text{Uniform}(0.05,0.5)$). In all simulations, $\delta_j \sim \text{Uniform}(0.01,0,082)$. This implies that, on average, the direct effect of each genetic variant on $X$ explains 0.1% of the variance in $X$.

The parameters $\rho_0 \in [0,1]$ and $\rho_\psi \in [0,1]$ control the proportions of variants with non-zero $\alpha_0$ and $\psi_j$, selected randomly and independently from the set of $L$ variants. For the direct effect of $G_j$ on $U_j$, $\psi_j = I_{\psi\,j}\psi_j'$, where $I_{\psi\,j} \sim \text{Bernoulli}(\rho_\psi)$ and $\psi_j' \sim \text{Uniform}(0.01,0,082)$. For the direct effect of $G_j$ on $Y_j$, $\alpha_{0\,j} = I_{0\,j}\alpha_{0\,j}'$, where $I_{0\,j} \sim \text{Bernoulli}(\rho_0)$ and $\alpha_{0\,j}' \sim \text{Uniform}(0.01,0,082)$.

### 5.2. Simulation scenarios

In all scenarios, $L \in \{50,100,150,200\}$ and $n_X = n_Y = \{25000n\}_{n=1}^{8}$. The remaining parameters differed for different scenarios and are described below.

### 5.2.1. Scenario 1: all variants are valid instruments – i.e., no horizontal pleiotropy

In the first scenario, there is no horizontal pleiotropy. For this, $\rho_0 = \rho_\psi = 0$ and $\beta \in \{0,0.5\}$.

### 5.2.2. Scenario 2: Directional horizontal pleiotropy; INSIDE holds

For this and the remaining scenarios, $\beta = 0$. In this scenario, there is directional (i.e., non-balanced) horizontal pleiotropy under the INSIDE assumption. For this, $\rho_\psi = 0$ and $\rho_0 = 1$.

### 5.2.3. Scenario 3: Directional horizontal pleiotropy; INSIDE is violated

In this scenario, there is directional horizontal pleiotropy that violates the INSIDE assumption. For this, $\rho_0 = 0$, $\rho_\psi \in \{0.5,0.75,1\}$.

## 5.3. Statistical analysis

In the simulations and applied example (section 6), we used the Koenker's studentized version[15] of the Breusch-Pagan test for heteroscedasticity.[16] We assessed two versions of the test: (i) the standard implementation, where the conditional variance model has the same structure as the mean model. That is, we assumed the following model for the conditional variance: $\text{Var}\left[\Gamma_j^*|\gamma_j^*\right] = \theta_0 + \theta_1\gamma_j^*$. (ii) a more flexible implementation, assuming the following model for the conditional variance: $\text{Var}\left[\Gamma_j^*|\gamma_j^*\right] = \varphi_0 + \varphi_1\gamma_j^* + \varphi_2{\gamma_j^*}^2$. Both versions of the test were applied for each of the following implementations of MR-Egger regression of $\hat{\Gamma}_j^*$ on $\hat{\gamma}_j^*$: (i) weighted using 1st order weights: $\sigma_{\Gamma_j}^{-2}$, which ignore uncertainty in $\hat{\gamma}_j$ (i.e., essentially assumes $\sigma_{\gamma_j}$ is negligible);[10] (ii) weighted using modified weights: $\left(\sigma_{\Gamma j}^2 + \hat{\beta}_{Egger}^2\sigma_{\gamma j}^2\right)^{-1}$, which incorporates uncertainty in $\hat{\gamma}_j$ as an approximation;[17] and (iii) unweighted: all genetic variants have the same weight (i.e., an unweighted simple linear regression of $\hat{\Gamma}_j^*$ on $\hat{\gamma}_j^*$). Since (i) is a typical implementation of MR-Egger regression, we considered it the primary implementation in our simulations. Implementations (ii) and (iii) were included to assess sensitivity to the choice of weighting scheme.

Even if INSIDE holds for the coding scheme that produces $\gamma_j^* > 0$ for all $j$, it may not hold for the coding scheme that produces $\hat{\gamma}_j^* > 0$ for all $j$. More specifically, even if $\text{Cov}\left(\text{sgn}(\gamma_j)|\gamma_j|, \text{sgn}(\gamma_j)\Gamma_j\right) = \text{Cov}\left(\gamma_j^*, \Gamma_j^*\right) = 0$, it is possible that $\text{Cov}\left(\text{sgn}(\hat{\gamma}_j)|\gamma_j|, \text{sgn}(\hat{\gamma}_j)\Gamma_j\right) \neq 0$. This is because the two coding schemes may be different. More specifically, it is possible that, due to sampling variability, $\text{sgn}(\gamma_j) \neq \text{sgn}(\hat{\gamma}_j)$, in which case the allele associated with positive values will differ.[9,18] To assess the influence of inducing INSIDE violations due to miscoding in the simulations, we considered two additional implementations of the Breusch-Pagan test: (iv) applying MR-Egger regression (1st order weights) to $\text{sgn}(\gamma_j)|\hat{\gamma}_j|$ and $\text{sgn}(\gamma_j)|\hat{\Gamma}_j|$ (instead of $\hat{\gamma}_j^*$ and $\hat{\Gamma}_j^*$), which corresponds to using estimated coefficients coded correctly (i.e., coded so that the effect allele choice results in $\gamma_j^* > 0$ for all $j$); (v) applying MR-Egger regression (1st order weights) to $\hat{\gamma}_j^*$ and $\hat{\Gamma}_j^*$, but excluding the weakest 10% of the instruments (instrument strength was quantified by $|\hat{\gamma}_j|/\sigma_{\gamma_j}$), which are the instruments at the greatest risk of miscoding.

Mean bias, coverage (i.e., the proportion of simulated datasets where the 95% confidence intervals included the true causal effect) and rejection rate (i.e., the proportion of simulated datasets where the 95% confidence intervals did not include the null) of MR-Egger regression slope were calculated over 10,000 simulations. Rejection rates of heteroscedasticity tests were also calculated. We also calculated the rejection rates of two heterogeneity tests: Cochran's $Q$ (which tests the null hypothesis that $\beta_{R_1} = \cdots = \beta_{R_L}$) and Rucker's $Q'$ (which tests the null hypothesis that $(\Gamma_1^* - c)/\gamma_1^* = \cdots = (\Gamma_L^* - c)/\gamma_L^*$, where $c$ is MR-Egger's intercept estimand).[10] We considered the following three implementations of these tests: (i) weighted using 1st order weights;[14] (ii) weighted using modified weights;[17] and (iii) weighted using 1st order weights and applying MR-Egger regression to $\gamma_j^*$ instead of $\hat{\gamma}_j^*$ (in which case 1st order weights are correct). Implementation (i) was considered the primary implementation, and implementations (ii) and (iii) were performed to assess sensitivity to weak instrument bias.

### 5.4. Results

Results for $L = 100$ and $n_X = n_Y = 100{,}000$ are shown in Table 1. For scenario 1 (no horizontal pleiotropy), when there was no causal effect, there was no bias, coverage was close to the expected 95%, and rejection rates were all close to or smaller than the expected 5%. For a non-null causal effect, there some bias towards the null (as expected in the two-sample setting) and coverage considerably lower than 95%. Moreover, the rejection rates of the heterogeneity tests were substantially inflated. The reduced coverage and inflated rejection rates are due to weak instrument bias, as we discuss below. Notably, the rejection rate of the heteroscedasticity test was virtually unaffected by the value of the causal effect.

In scenario 2 (horizontal pleiotropy, but INSIDE holds), MR-Egger had virtually no bias and coverage close to 95% (it gets even closer for larger sample sizes, as discussed below). While heterogeneity tests had rejection rates close to 100%, the figures for heteroscedasticity tests where smaller than 10%. Although this broadly corroborates the discussion in section 4.6, the rejection rates for the heteroscedasticity tests were slightly inflated, and we discuss why below. In scenario 3 (horizontal pleiotropy violating INSIDE), MR-Egger presented bias and undercoverage. The heteroscedasticity tests presented rejection rates considerably larger then in scenario 2, corroborating the test is specific to (at least some forms of) INSIDE violation. Increasing the proportion of invalid instruments reduced, rather than increased, the power to detect INSIDE violations. Although this may seem counter-intuitive, it is expected. This happens because, in the data-generating model used in the simulations, increasing $\rho_\psi$ reduces the variability in $\psi_j k_{Y_j}$ by reducing the proportion of $\psi_j$ terms that are equal to zero. Even though such behaviour is not necessarily expected in practice (for example, if different invalid instruments violate INSIDE by affecting different confounders, then increasing the number of invalid instruments would be expected to increase the variability in the $\psi_j k_{Y_j}$ terms), this example reinforces that proposed testing procedure should be viewed as a falsification, but not confirmatory, test of the plausibility of the INSIDE assumption.

In Supplementary Figure 1, we explore weak instrument bias in (scenario 1, non-null causal effect) in more detail. For both heterogeneity tests (panels A and B), the rejection rates were considerably inflated, especially for larger numbers of variants, and there was no clear indication of convergence towards the expected 5% for the sample sizes considered. This is due to weak instrument bias, as the results using the true (rather than estimated) $G_j \rightarrow X$ coefficients were at the expected 5%. Moreover, the rejection rates using the modified weighting scheme overlapped nearly perfectly with the latter. However, with respect to MR-Egger regression slope, both bias (panel C) and coverage (panel D) were similar between the two weighting schemes, and at the expected values (0 and 95%, respectively) when using true coefficients.

Figure 2 displays the rejection rate of the heteroscedasticity test for patterns of horizontal pleiotropy (panels A-F), sample size (horizontal axis) and number of genetic instruments (colours). In the absence of horizontal pleiotropy (panels A and B), the rejection rates were close to 5%. In the presence of horizontal pleiotropy, but no asymptotic INSIDE violation (panel C), the rejection rates converge to 5% as sample size increases. As shown in panels D-F, the rejection rate is substantially higher than 5% in the presence of INSIDE violation, and power increases with both number of instruments and sample size. In general, the power to detect INSIDE violations was greater for the linear specification. The results were virtually unchanged when using modified weights (Supplementary Figure 3) or no weights (Supplementary Figure 4).

In panel C in Figure 3, the rejection rates were clearly inflated, mainly for smaller sample sizes and more noticeably for the quadratic specification. This is an important issue which could indicate that the test is reliable only in very large samples. However, in this scenario the rejection rate of the heteroscedasticity test varied linearly with the rejection rate of MR-Egger regression and, given $L$, with the bias of MR-Egger causal effect estimate (Supplementary Figure 4). This indicates that the test was not over-rejecting, but rather detecting sample violations of INSIDE. More specifically, this happens because, even though in the simulations $\gamma_j > 0$ for all $j$, it is possible that $\hat{\gamma}_j < 0$ for some $j$, specifically for the weaker genetic instruments. Such variants are recoded (since the all-positive recoding step performed by MR-Egger uses estimated coefficients), which creates the possibility for sample INSIDE violations, since in the simulations INSIDE holds for the coding scheme corresponding to positive values of the true, rather than estimated, coefficients. This can be seen in Supplementary Figure 5: using $\hat{\gamma}_j$ (positive coding with respect to true $G_j \rightarrow X$ coefficients coefficients) rather than $\hat{\gamma}_j^*$ (positive coding with respect to estimated $G_j \rightarrow X$ coefficients coefficients) eliminated any inflation in the rejection rate. Such elimination was also produced by excluding the weakest 10% of the instruments.

## 6. Applied example

We illustrate the practical application of the proposed approach by re-analysing a real dataset of summary genetic association results involving XL-HDL-C as the exposure (genetic associations are expressed as per-allele standard deviation changes) and AMD as the outcome (genetic associations are expressed as per-allele ln(odds ratio)). The dataset contains 27 genetic variants selected from three studies,[19-21] as described elsewhere[22] and available in Supplementary Table. A scatter plot of the dataset is shown in Figure 3, from which a single genetic variant (marked in red) can be clearly identified as an outlier.

In addition to MR-Egger regression, we considered three MR methods: the inverse-variance weighted (IVW) estimator,[23] which is the baseline two-sample MR method[3] in the sense it requires all variants to be valid instruments (or that there are positive and negative biases that cancel out); simple median, which is consistent if more than 50% of the variants are valid instruments; and the weighted median,[24] which is consistent if more than 50% of the weights come from variants that are valid instruments. Median estimators were chosen for their insensitivity to outliers, which is relevant in the present analysis given the clear outlier (Figure 3). Simple and weighted versions of the median method were considered because the outlier had the largest value of $\hat{\gamma}_j^*$, and therefore has a large influence on weighted estimators.

As shown in Table 2, when all 27 variants are included in the analysis, point estimates from different estimators were inconsistent: IVW was close to null, simple median suggested a positive (i.e., risk-increasing) effect while weighted median and MR-Egger regression suggested a negative effect. The similarity between weighted median and MR-Egger is expected given that both are strongly influenced by the outlying variant, which has the largest value of $\hat{\gamma}_j^*$ (and thus a large influence on weighted estimators) and a negative ratio estimate of -0.80. However, for the simple median, the outlier has the same weight as any other variant and the estimator is therefore less influenced by it compared to weighted estimators. Moreover, there is strong statistical evidence against the null hypothesis of homogeneity from both heterogeneity tests. Finally, the heteroscedasticity test provided strong statistical evidence against the null hypothesis, thus suggesting that INSIDE is violated.

Applying the heteroscedasticity test in a leave-one-out procedure revealed that removing the single outlying variant virtually eliminated statistical evidence for heteroscedasticity, as measured using both the P-value of the Breusch-Pagan test and the Pearson correlation coefficient between the absolute value of the residuals from MR-Egger regression and $\hat{\gamma}_j^*$ (Table 2). Of note, this was not the case for any other variant. Therefore, the subset of 26 variants that excludes the outlier is the largest subset where homoscedasticity holds. Under the assumption described in section 4.5, INSIDE holds in this subset, and MR-Egger regression yields a consistent estimate of the causal effect. Indeed, after excluding the single outlier, results from all estimators were reassuringly similar and close to the original estimates produced by the simple median, which this estimator is the least influenced by the outlier. Moreover, after excluding the outlying variant, both heterogeneity tests still reject (although less strongly) the null hypothesis. In combination with the results for the entire set of 27 variants, this corroborates the notion that, in this context, heteroscedasticity implies heterogeneity, but not the other way round (section 4.6).

As discussed in section 4.4.1, the fact that $\hat{\Gamma}_j$ is expressed as ln(odds ratio) may lead to heteroscedasticity in the absence of INSIDE violation. Indeed, the Pearson correlation coefficient between $\hat{\sigma}_{\Gamma j}$ and $\hat{\gamma}_j^*$ was 0.43 (P=0.026). We assessed this possibility using simulations where $\hat{\Gamma}_j' \sim N\left(0, \hat{\sigma}_{\Gamma j}^2\right)$ (so that the correlation between $\hat{\sigma}_{\Gamma j}$ and $\hat{\gamma}_j$ becomes the only possible source of heteroscedasticity) and applied the heteroscedasticity test using $\hat{\Gamma}_j'$ instead of $\hat{\Gamma}_j$. Over 10,000 simulations, the mean P-value for the test was 0.53, with a rejection rate of 3%. Therefore, the correlation between $\hat{\sigma}_{\Gamma j}$ and $\hat{\gamma}_j^*$ is unlikely to explain the results.

## 7. Discussion

In this paper we showed that the coding scheme used in MR-Egger regression is equivalent to a model that is coding invariant – that is, each term in the regression equation is invariant under allele recoding. Moreover, we described how the all-positive coding scheme can be interpretated in relation to the bias in ratio estimands produced by direct effects of the candidate genetic instruments on the outcome. This of course does not mean that the all-positive coding version of INSIDE is (or not) to be preferred over the VICE assumption, much less that is a plausible assumption at all; rather, this suggests that MR-GRIP does not seem to be necessarily superior to MR-Egger regression with respect to coding dependency or lack thereof; and (ii) it seems at least in principle possible to use substantive knowledge to assess the plausibility of the INSIDE assumption. For example, if some biological mechanisms are believed to bias ratio estimands, one can judge whether such mechanisms are also plausibly related to instrument strength (e.g., if such mechanisms mediate the effect of at least some of the candidate genetic instruments on the exposure). If so, then INSIDE would be deemed implausible. Importantly, this (admittedly simplistic) reasoning relies on biological understanding without making any reference to coding scheme. However, it should be noted that, even if one is satisfied with INSIDE with respect to GRIP and interpretability, this does not eliminate other potential difficulties associated with MR-Egger regression, such as low power partly due to a coding scheme that reduces variability in the coefficients relating the genetic variants and the exposure.[8,9]

Both in the case of INSIDE and VICE, the independence (or at least uncorrelatedness) assumption involves quantities that include $\gamma_j$ (or some function of it). In the case of VICE, $\gamma_j$ itself is a factor in both $\alpha_j \gamma_j$ and $\gamma_j^2$; for MR-Egger, $\text{sgn}(\gamma_j)$ is a factor in both $\alpha_j \mathrm{sgn}(\gamma_j)$ and $|\gamma_j| = \mathrm{sgn}(\gamma_j)\gamma_j$. At least from a statistical perspective, this indicates that the two quantities relevant to each assumption are not expected to be independent in general. This reinforces the importance of carefully assessing the plausibility of these assumptions, including empirical tests. In this respect, we propose a simple way to assess the plausibility of these assumptions in two-sample MR by applying a heteroscedasticity test to a summary dataset of genetic associations. We also proposed an additional assumption – that INSIDE holds (rather than being identically violated) in the largest homoscedastic subset of variants – that allows correcting for bias driven by INSIDE violation by identifying the largest subsets of variants where homoscedasticity holds. Our simulation study corroborated the theoretical findings, and we illustrated how the method can be applied in practice by re-analysing a real dataset.

The fact that removing only a specific variant virtually eliminated heteroscedasticity in the applied example greatly simplified the process of identifying the largest subset of variants where homoscedasticity holds. This allowed illustrating how heteroscedasticity tests can be used in practice to mitigate bias under the assumption that INSIDE holds in the largest homoscedastic subset of variants. However, in many cases it may be necessary to search over many more subsets, thus rendering an exhaustive search unfeasible when $L$ is relatively large. For example, if $L = 20$, there are 1,048,365 unique subsets of variants containing at least three variants each. For $L = 30$, there are 1,073,741,358 such combinations. Although it would be tractable to explore, for example, the 30 subsets containing 29 variants, then the 435 subsets containing 28 variants etc. for a few iterations, the process quickly becomes intractable: for example, there are 30,045,015 subsets containing 20 variants. We acknowledge that the development of efficient algorithms that exploit the same concept in a computationally tractable way is required for the correction method to be more widely applicable. In this regard, a recent paper described how Bayesian model averaging can be used to optimize the search through subsets of variants to efficiently identify subsets of homogeneous variants.[22] The same study also showed how the method be extended for a two-parameter model (in this case, two slopes and no intercept), thus allowing consistent causal effect estimation even if some variants violate the INSIDE assumption. Indeed, if two slopes are identified in the data, this would by itself be evidence against INSIDE under the model described in equation (3). Otherwise, multiple slopes could also be present due to causal effect heterogeneity, and it would not be possible to distinguish between bias and "true" heterogeneity from the data alone.

Even though the Bayesian model averaging approach has important advantages over the approach based on heteroscedasticity testing (i.e., efficient search of subsets of variants and possibility of identifying two slopes), it also has the important limitation of not being able to identify sets of homogeneous variants in situations where INSIDE holds. For example, if all variants are non-identically invalid but homoscedasticity holds, then all ratio estimands would be different and, asymptotically, no two variants would be clustered together. However, the heteroscedasticity test would not reject, thus indicating that INSIDE holds (or, more specifically, not falsifying INSIDE) without any filtering process. The most fruitful strategy is likely to use both approaches and compare the results between them. Moreover, it may be possible to combine the methods by adapting the Bayesian model averaging approach to identify subsets of homoscedastic rather than homogeneous variants, but this possibility remains to be formally investigated.

Our main goal was to demonstrate the connection between INSIDE and heteroscedasticity, thus providing a theoretical justification for the use of heteroscedasticity tests to assess the plausibility of the INSIDE assumption. In the simulation study and applied example, we used the Breusch-Pagan test (described in section 5.3). Although the results obtained using used the standard implementation of the test aligned with the theoretical expectation, it is in principle possible that the model is mis-specified, which could for example reduce statistical power. In the simulations, we considered a linear and a quadratic specification, and none was clearly superior to the other. However, this does not imply this will always be the case, and of course additional specifications could be considered. In the absence of a biologically motivated data-generating process that implies a specific model for the conditional variance (which would likely be context-specific), opting for simpler specifications that are likely to detect heteroscedasticity at least to some degree (possibly allowing for some non-linearity) while avoiding an overly complex model seemed a reasonable pragmatic compromise. Of course, in addition to model specification within a single test, there are alternative testing procedures that could be considered. Such comparison across different tests and implementations are beyond the scope of this paper and require future methodological work.

The simulations also highlighted the issue of weak instrument bias. In the case where all variants were valid instruments and there was a non-null causal effect, the rejection rates of heterogeneity tests using 1st order weights were considerably inflated and presented no indication of convergence to the expected 5% within the sample sizes we assessed. However, modified weights presented much better results, suggesting this weighting strategy is preferred to mitigate bias in this setting. However, MR-Egger regression was influenced by weak instrument bias regardless of the weighting scheme: even though bias and coverage presented some indication of convergence to zero and 95%, respectively, they were still considerably far from these values within the sample sizes we assessed. Other approaches, such as simulation extrapolation[25] and bootstrapping, can be used, but we have not explored the issue further as this was not the focus of the present paper. Notably, the results of the Breusch-Pagan test depended very little on the weighting scheme. Another bias source in MR-Egger regression that relates to instrument strength is when the signs of estimated $G_j \to X$ coefficients differ from the signs of the true coefficients.[9,18] In our simulations, this resulted in INSIDE violations even when this assumption held with respect to the all-positive (for the true coefficients) coding scheme. Therefore, in MR-Egger regression, there is this second form of weak instrument bias that can occur even when there is a null causal effect. This was attenuated by removing the weakest 10% of the instruments, as expected. Although the specific cutoff of 10% is unlikely to generalise to all settings (it was a rather arbitrary choice to illustrate a point in the simulations), this result suggests that it may be a useful sensitivity analysis to remove a subset of the weakest instruments to mitigate potential bias due to positive allele miscoding. This has analogies to methods such as MR-CORGE,[26] which postulates stronger instruments are more likely to be valid (i.e., less likely to have direct effects on the outcome) then weaker instruments. Our point is not about true direct effects of genetic variants, but the higher risk of weaker variants to suffer from positive allele miscoding which could in turn lead to INSIDE violations.

Similar to using heterogeneity tests to assess horizontal pleiotropy, heteroscedasticity tests should be used as one of the tools for exploring the plausibility of the INSIDE assumption, rather than as the single definitive criterion for declaring the assumption as valid or invalid in a particular situation. INSIDE is not fully empirically verifiable. Therefore, as discussed in section 4.4, it is possible that the test rejects when INSIDE holds and does not reject when INSIDE is violated. This and other critical assumptions for consistent causal effect estimation using MR should be assessed using multiple complementary methods,[3] such as multivariable MR[27] (which allows adjusting for one or more potential confounders affected by one or more instruments), gene×covariable interactions[28] (which exploits variability in instrument strength to quantify and correct for horizontal pleiotropy, including scenarios where INSIDE is violated) and causal effect estimators that make different identifying assumptions[3] (where consistency between methods would strengthen causal conclusions). Indeed, even heteroscedasticity itself can be exploited in ways other than those proposed here in IV analysis, such as for consistent causal effect estimation in methods such as MR-GENIUS;[29] and for empirically assessing homogeneity in the association between $G$ and $X$.[30] All methods require their own assumptions, but coherent results across a range of methods that make different assumptions increases the plausibility that the conclusions are robust and not mainly driven by bias.

**Acknowledgements**

FPH and GDS work within the Medical Research Council (MRC) Integrative Epidemiology Unit at the University of Bristol, which is supported by the MRC (grant MC_UU_00032/1). JB is funded by the MRC (grant MR/X011372/1). JB and FB are funded by the MRC (grant MC/MR/WO14548/1). FPH is supported by a research productivity fellowship from the Brazilian National Council for Scientific and Technological Development (grant 303880/2023-6). This is a summary of independent research carried out at the National Institute for Health and Care Research (NIHR) Leicester Biomedical Research Centre (BRC). The views expressed are those of the authors and not necessarily those of the NIHR or the Department of Health and Social Care.

This work is dedicated to the memory of Franciso(a) (FPH's unborn child).

**Data availability statement**

The data used in this article are available as supplementary material to this article. The R code used for simulations and empirical data analysis is available at: https://github.com/FernandoHartwig/INSIDE_test.

**TABLES**

**Table 1.** Mean bias, coverage (in %) and rejection rate (RR, in %) of MR-Egger regression slope, and rejection rates of the heterogeneity and heteroscedasticity tests, for $L = 100$ and $n_X = n_Y = 100{,}000$.

| Statistic | Scenario 1 | | Scenario 2 | Scenario 3 | | |
|---|---|---|---|---|---|---|
| | $\boldsymbol{\beta = 0}$ | $\boldsymbol{\beta = 0.5}$ | | $\boldsymbol{\rho_\psi = 0.5}$ | $\boldsymbol{\rho_\psi = 0.75}$ | $\boldsymbol{\rho_\psi = 1}$ |
| Bias[A] | 0.00 | -0.03 | 0.01 | 0.14 | 0.16 | 0.14 |
| Coverage[A] | 95.5 | 82.4 | 93.9 | 8.7 | 4.8 | 8.5 |
| RR[A] | 4.5 | 100 | 6.1 | 91.3 | 95.2 | 91.5 |
| RR[B] | 5.1 | 48.7 | >99.9 | >99.9 | >99.9 | >99.9 |
| RR[C] | 4.9 | 45.7 | >99.9 | >99.9 | >99.9 | >99.9 |
| RR[D] | 2.8 | 2.8 | 6.8 | 56.0 | 33.0 | 23.2 |
| RR[E] | 2.8 | 2.5 | 9.0 | 43.9 | 28.3 | 19.4 |

[A]MR-Egger regression slope. [B]Heterogeneity (Cochran's Q) test [C]Heterogeneity (Rucker's Q') test. [D]Heteroscedasticity (Breusch-Pagan) test, linear specification. [E]Heteroscedasticity (Breusch-Pagan) test, quadratic specification.

**Table 2.** Causal effect estimates of large particle high density lipoprotein cholesterol (XL-HDL-C) on age-related macular degeneration (AMD), expressed as ln(odds ratio of AMD) per standard deviation increase in XL-HDL-C, and results from heterogeneity and heteroscedasticity tests in the applied example.

| **Method** | **Before outlier removal** | **After outlier removal** |
|---|---|---|
| *Causal effect estimates (95% confidence intervals)* | | |
| IVW | 0.03 (-0.35 to 0.40) | 0.65 (0.33 to 0.97) |
| Simple median | 0.69 (0.00 to 1.37) | 0.72 (0.08 to 1.35) |
| Weighted median | -0.11 (-0.71 to 0.50) | 0.68 (0.36 to 1.00) |
| MR-Egger regression | -0.13 (-0.60 to 0.33) | 0.70 (0.27 to 1.13) |
| *Heterogeneity tests: Q statistic (degrees of freedom) and P-values* | | |
| Cochran's Q | 115.96 (df=26); P<0.0001 | 40.62 (df=25); P=0.0252 |
| Rucker's Q' | 109.38 (df=25); P<0.0001 | 39.69 (df=24); P=0.0231 |
| *Breusch-Pagan test (Koenker's studentized version)* | | |
| P-value | P<0.0001 | P=0.9564 |
| *MR-Egger regression residuals (absolute value) and* $\hat{\gamma}_j^*$ | | |
| Pearson correlation | 0.78 | 0.06 |

**FIGURES**

**Figure 1. Path diagram depicting causal relationships between the $j$th genetic instrument ($G_j$), the exposure ($X$), the outcome ($Y$) and an unmeasured variable that is potentially affected by $G_j$ ($U_j$, panel A) or that potentially affects $G_j$ ($W_j$, panel B).**

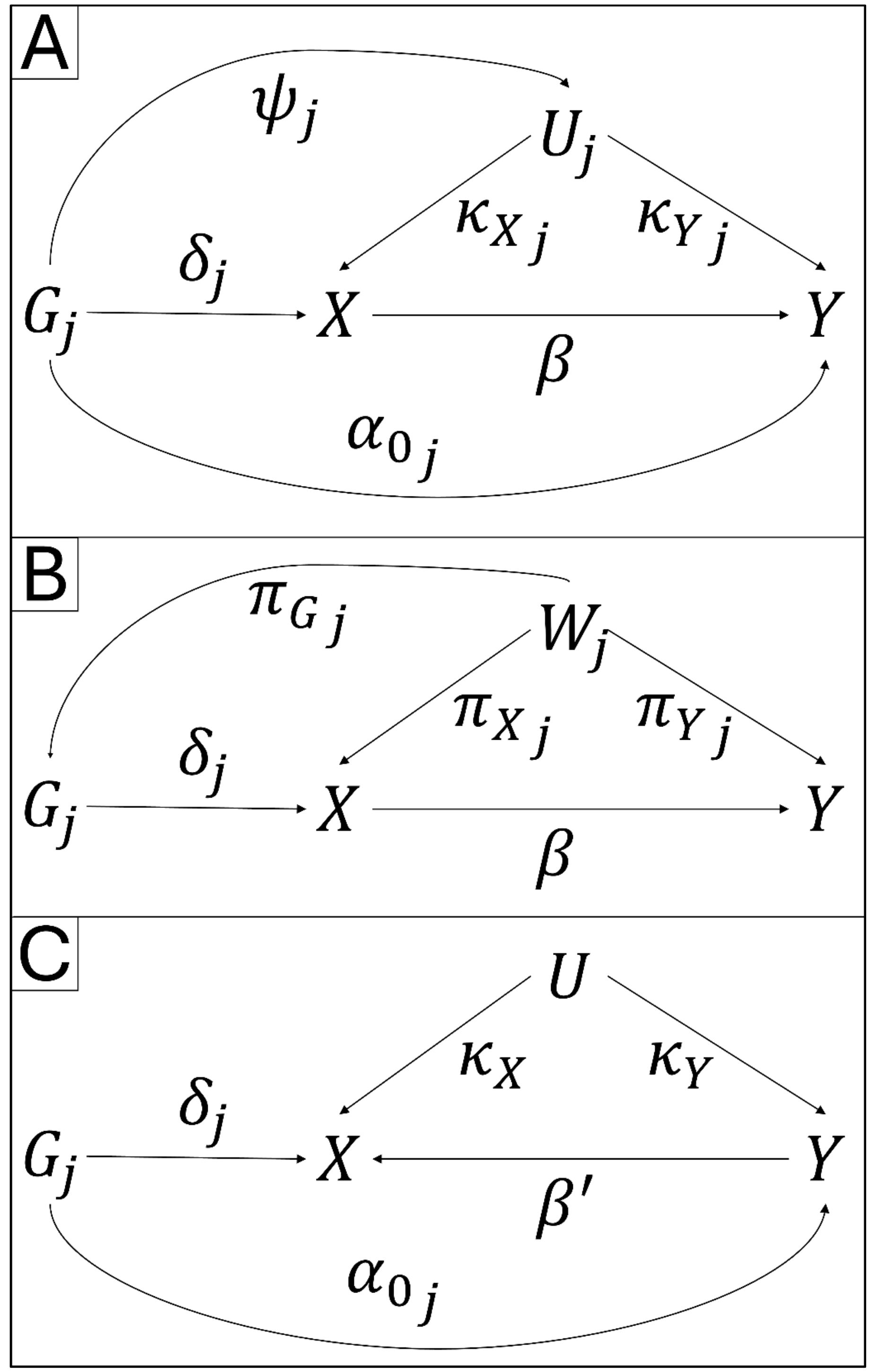

**Figure 2. Scatter plots of rejection rates of the Breusch-Pagan heteroscedasticity test (vertical axis) against sample size (horizontal axis), for $L = 50$ (black), $L = 100$ (grey), $L = 150$ (red) and $L = 200$ (blue). Circles and crosses respectively denote the linear and quadratic specifications.**

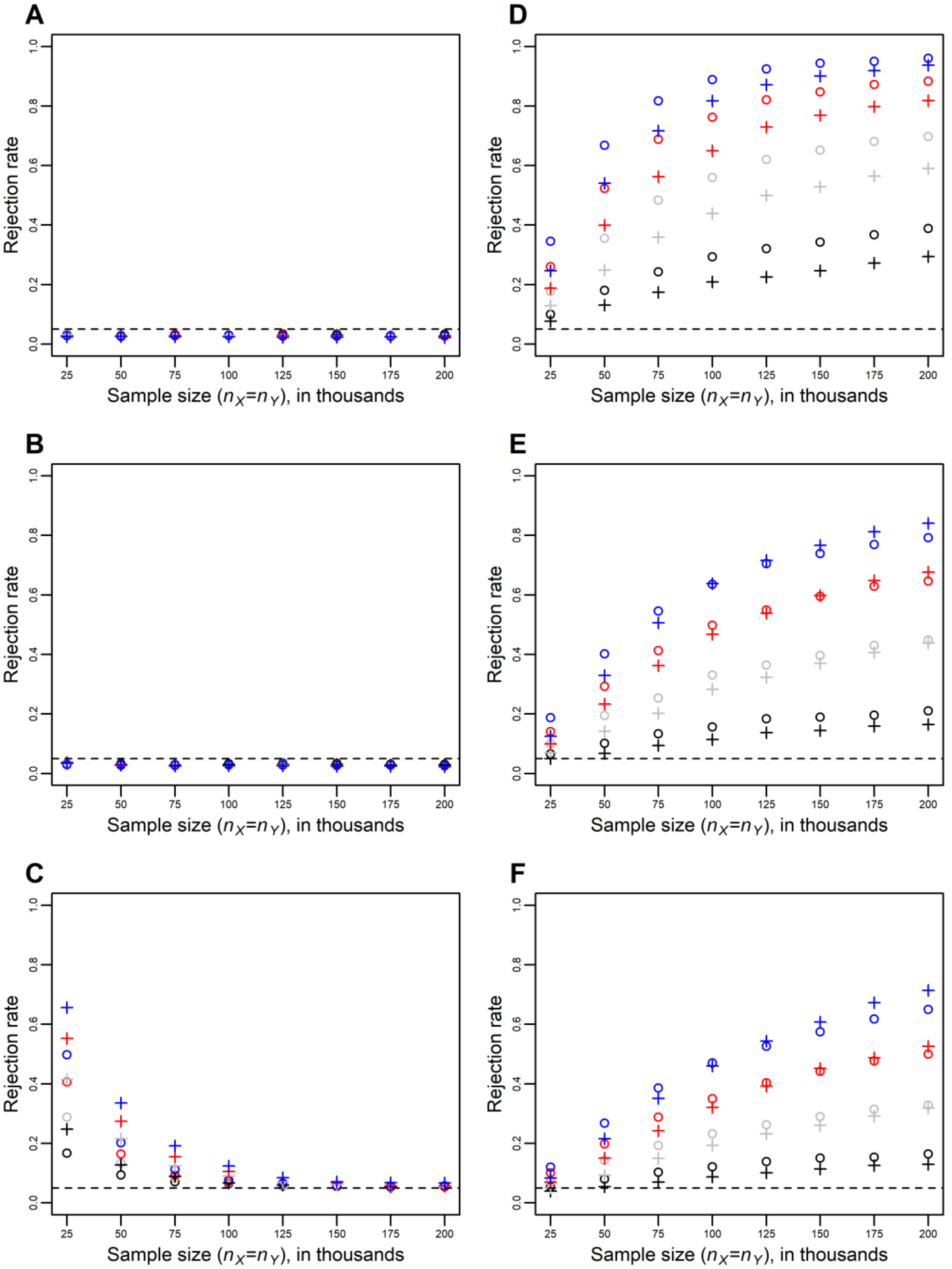

A,B: scenario 1, with $\beta = 0$ (A) and $\beta = 0.5$ (B). C: scenario 2. D-E: scenario 3, with $\rho_\psi = 0.5$ (D), $\rho_\psi = 0.75$ (E) and $\rho_\psi = 1$ (F).

**Figure 3. Scatter plot of summary associations of each genetic instrument with large particle high density lipoprotein cholesterol (XL-LDL-C)[a] and age-related macular degeneration (AMD)[b]. Dotted lines denote 95% confidence intervals. An outlier is marked in red.**

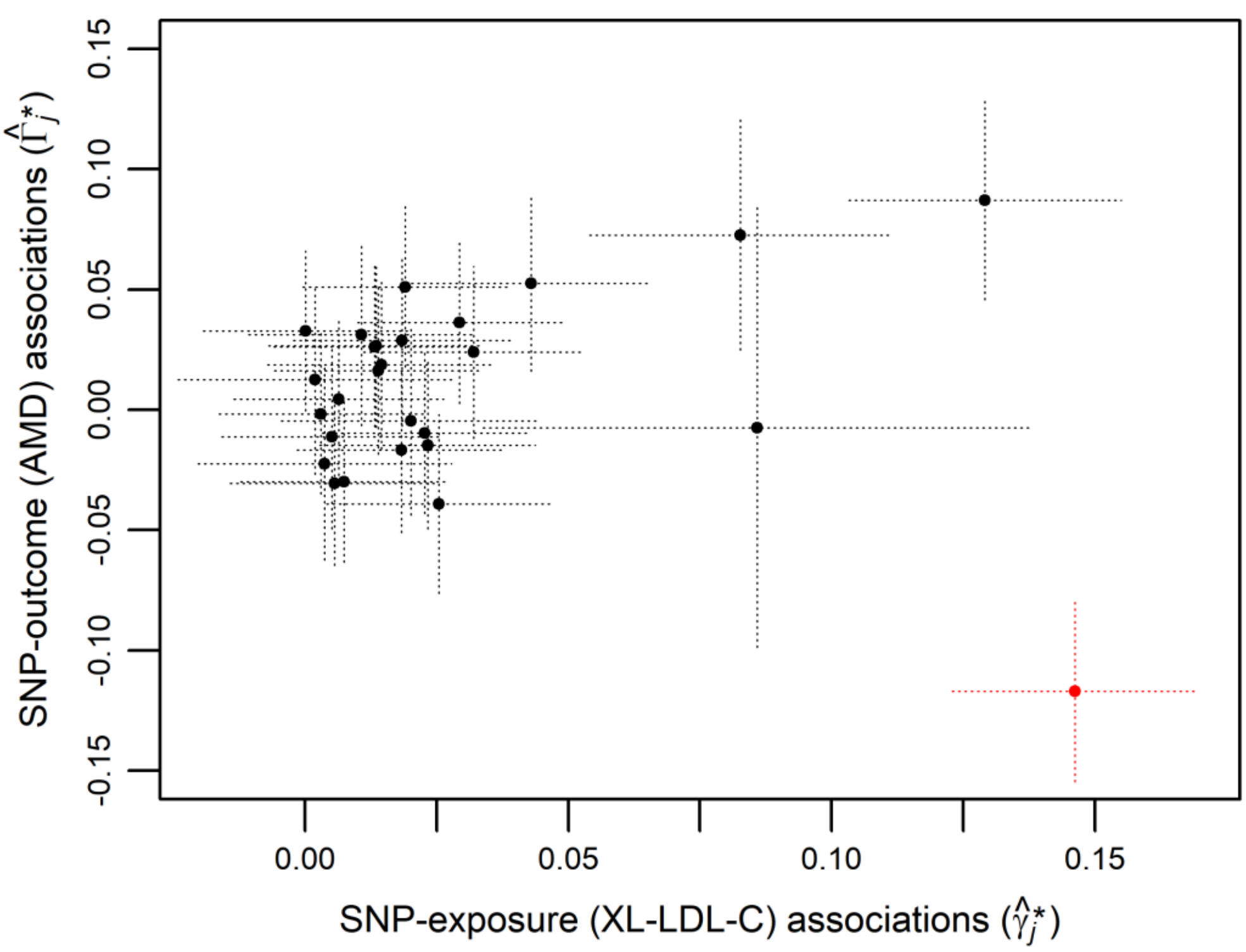

[a]Coefficients are per-allele standard deviation changes.

[b]Coefficients are per-allele log(odds ratio).

## SUPPLEMENTARY MATERIAL

**Supplementary Table 1. Summary associations of each genetic instrument with large particle high density lipoprotein cholesterol (XL-LDL-C), expressed as per-allele standard deviation changes and associated standard errors; and age-related macular degeneration (AMD), expressed as per-allele log(odds ratio) and associated standard errors.**

| XL-HDL-C | | AMD | |
|---|---|---|---|
| **Coefficient** | **Standard error** | **Coefficient** | **Standard error** |
| -0.025437 | 0.0109058752523311 | 0.039061434 | 0.018971625 |
| -0.018948 | 0.00990473801818187 | -0.05105715 | 0.017006347 |
| -0.005568 | 0.0101100626900669 | 0.030480926 | 0.017479946 |
| -0.018265 | 0.0100940075387559 | 0.016713328 | 0.017479946 |
| 0.013201 | 0.00974361892243782 | 0.026260264 | 0.016981183 |
| -0.006377 | 0.0101481937236723 | -0.004383264 | 0.017479946 |
| 0.001853 | 0.0132514088102056 | 0.012603113 | 0.020068459 |
| 0.029343 | 0.00989380026091665 | 0.036344293 | 0.017162108 |
| 0.003003 | 0.00988764327091527 | -0.001684587 | 0.01698615 |
| -0.010659 | 0.0109264037436127 | -0.031129342 | 0.019202815 |
| 0.01345 | 0.0104229427762118 | 0.0266903 | 0.017351658 |
| -0.013841 | 0.0100460781137884 | -0.016270195 | 0.017774541 |
| -0.00505 | 0.010660444836611 | 0.011253419 | 0.019456412 |
| 0.082639 | 0.0145875697313465 | 0.072687568 | 0.024380926 |
| 4.4e-05 | 0.00991437600459322 | 0.032750955 | 0.016983104 |
| 0.031978 | 0.0106801177001762 | 0.023921067 | 0.018211102 |
| 0.023254 | 0.0104639311019997 | -0.01468893 | 0.017714911 |
| 0.042871 | 0.0115919058551226 | 0.052513228 | 0.018655782 |
| -0.003729 | 0.0123066182947926 | 0.022498785 | 0.020354106 |
| 0.020106 | 0.0125813060797498 | -0.004591056 | 0.019950869 |
| 0.146223 | 0.0118766965727937 | -0.116997468 | 0.019198231 |
| 0.00738 | 0.00998506637424081 | -0.029836604 | 0.017071024 |
| 0.018377 | 0.0105161782811878 | 0.028819772 | 0.018133996 |
| 0.014442 | 0.0110223668674884 | 0.018785158 | 0.017901076 |
| -0.022716 | 0.00991873016175733 | 0.009732605 | 0.017044858 |
| 0.085882 | 0.0265603939345484 | -0.007397638 | 0.046649327 |
| 0.129142 | 0.0132168986973217 | 0.087128107 | 0.021154225 |

**Supplementary Figure 1. Scatter plots of the rejection rate of Cochran's Q (panel A) and Rucker's Q' (panel B) heterogeneity tests, and bias (panel C) and coverage (panel D) of MR-Egger regression slope against sample size in scenario 1 with a non-null causal effect. Black, grey, red and blue respectively denote, $L = 50$, $L = 100$, $L = 150$ and $L = 200$. Circles, cross-marks and crosses respectively denote first-order weights, modified weights and first-order weights using the true (rather than estimated) $G_j \rightarrow X$ coefficients.**

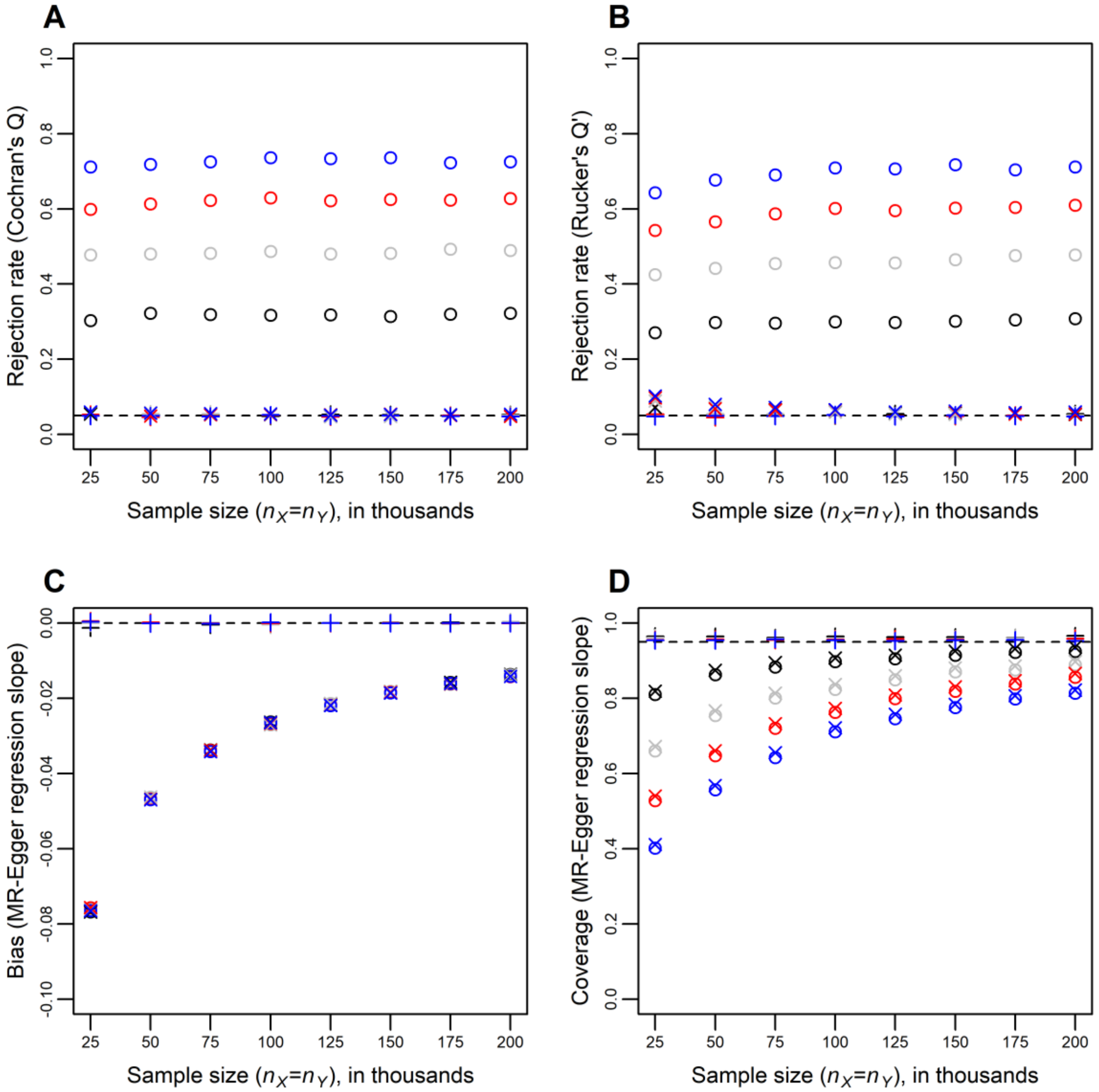

**Supplementary Figure 2. Scatter plots of rejection rates of the Breusch-Pagan heteroscedasticity test using modified weights (vertical axis) against sample size (horizontal axis), for $L = 50$ (black), $L = 100$ (grey), $L = 150$ (red) and $L = 200$ (blue). Circles and crosses respectively denote the linear and quadratic specifications.**

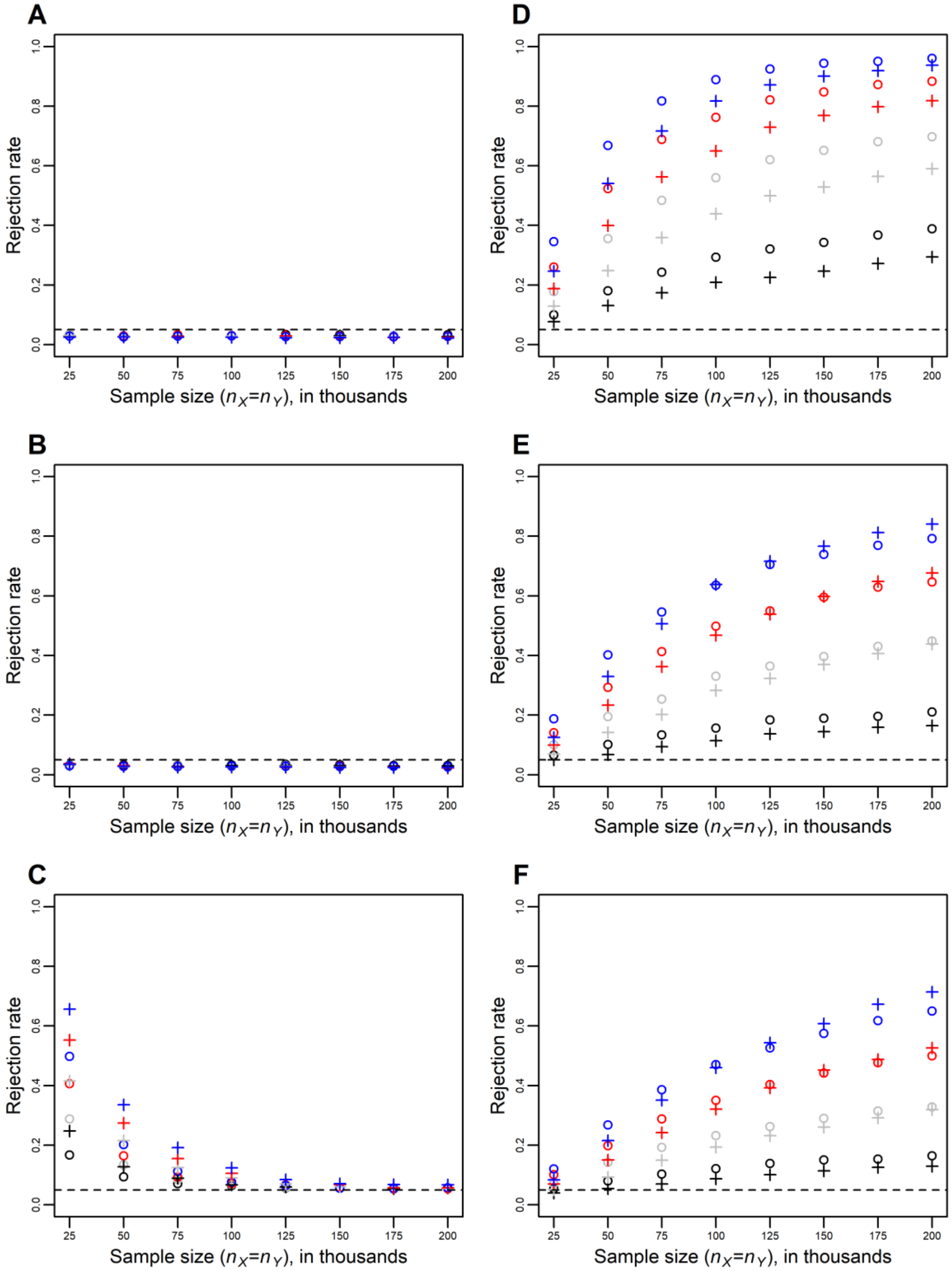

A,B: scenario 1, with $\beta = 0$ (A) and $\beta = 0.5$ (B). C: scenario 2. D-E: scenario 3, with $\rho_\psi = 0.5$ (D), $\rho_\psi = 0.75$ (E) and $\rho_\psi = 1$ (F).

**Supplementary Figure 3. Scatter plots of rejection rates of the unweighted Breusch-Pagan heteroscedasticity test (vertical axis) against sample size (horizontal axis), for $L = 50$ (black), $L = 100$ (grey), $L = 150$ (red) and $L = 200$ (blue). Circles and crosses respectively denote the linear and quadratic specifications.**

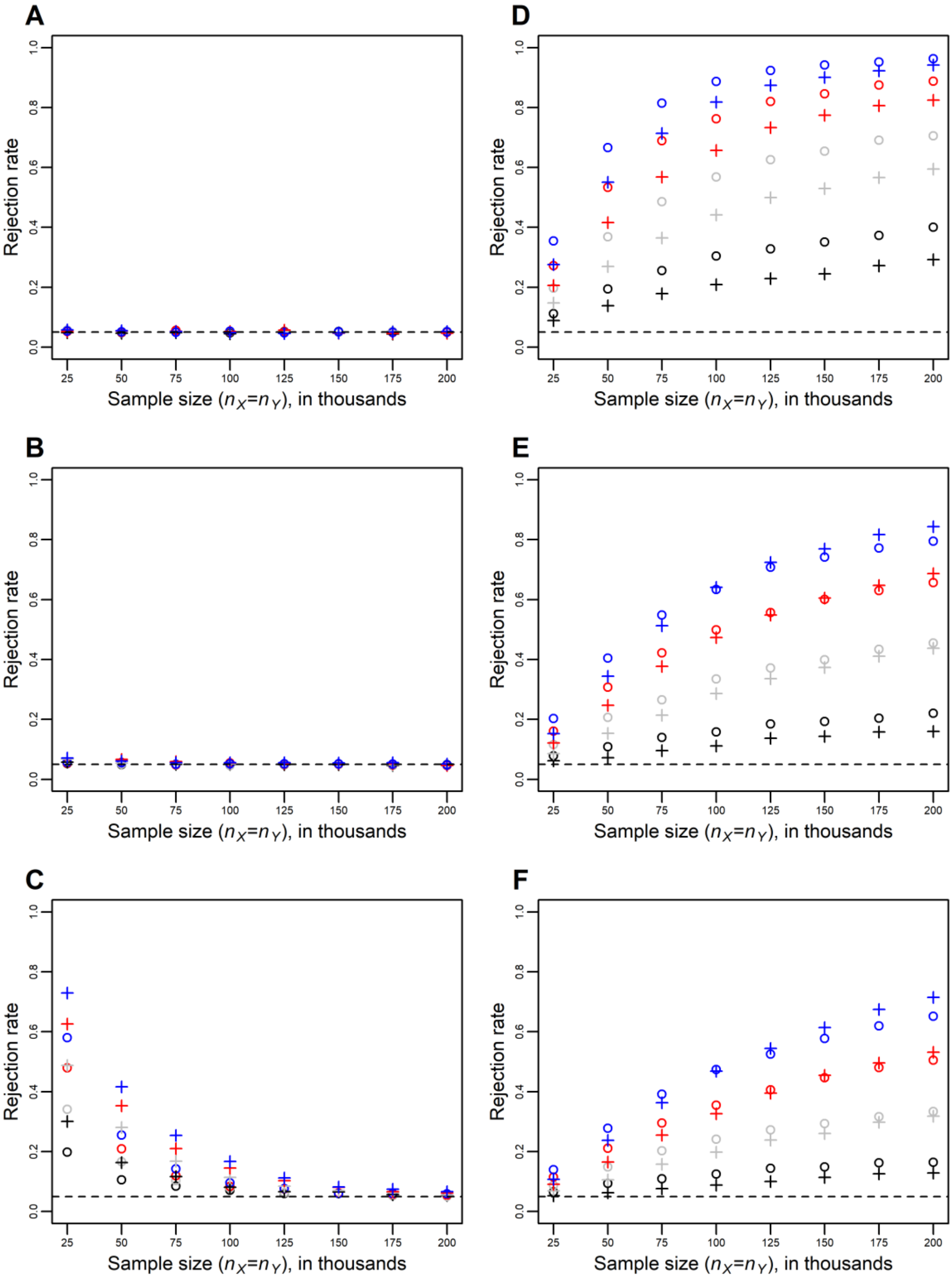

A,B: scenario 1, with $\beta = 0$ (A) and $\beta = 0.5$ (B). C: scenario 2. D-E: scenario 3, with $\rho_\psi = 0.5$ (D), $\rho_\psi = 0.75$ (E) and $\rho_\psi = 1$ (F).

**Supplementary Figure 4. Scatter plots of the rejection rate (RR) of the linear specification of the heteroscedasticity test (vertical axis) in scenario 2 of the simulation study against: RR of MR-Egger regression (right panel; dashed lines indicate RR=5%); and bias of the MR-Egger regression causal effect estimate (left panel; black, grey, red and blue respectively denote, $L = 50$, $L = 100$, $L = 150$ and $L = 200$).**

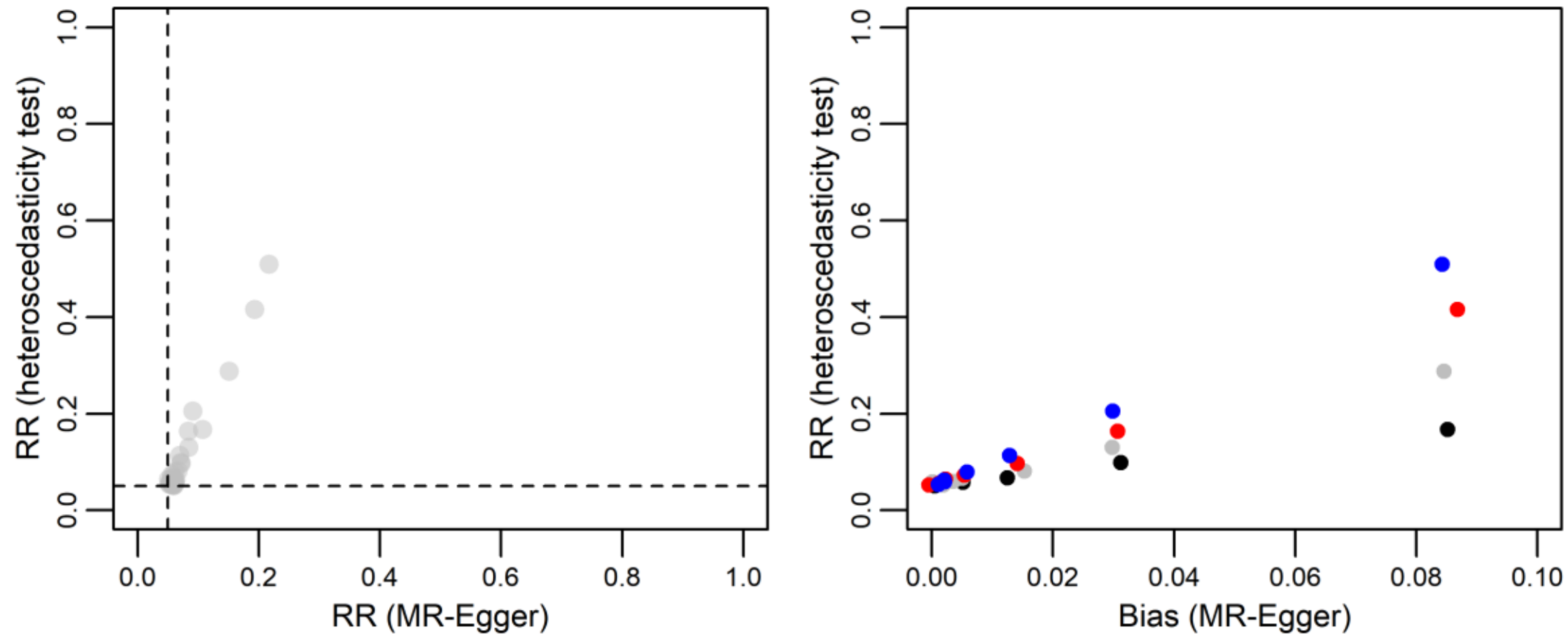

**Supplementary Figure 5. Scatter plots of the rejection rate of the linear (panel A) and quadratic (panel B) specifications of the Breusch-Pagan heteroscedasticity test, and bias (panel C) and rejection rate (panel D) of MR-Egger regression slope against sample size in scenario 2. Black, grey, red and blue respectively denote, $L = 50$, $L = 100$, $L = 150$ and $L = 200$. Circles, crosses and cross-marks respectively denote (i) allele coding so that estimated $G_j \rightarrow X$ coefficients are all positive, (ii) allele coding so that true coefficients are positive, (iii) allele coding so that estimated $G_j \rightarrow X$ coefficients are all positive and excluding the weakest 10% of the genetic instruments.**

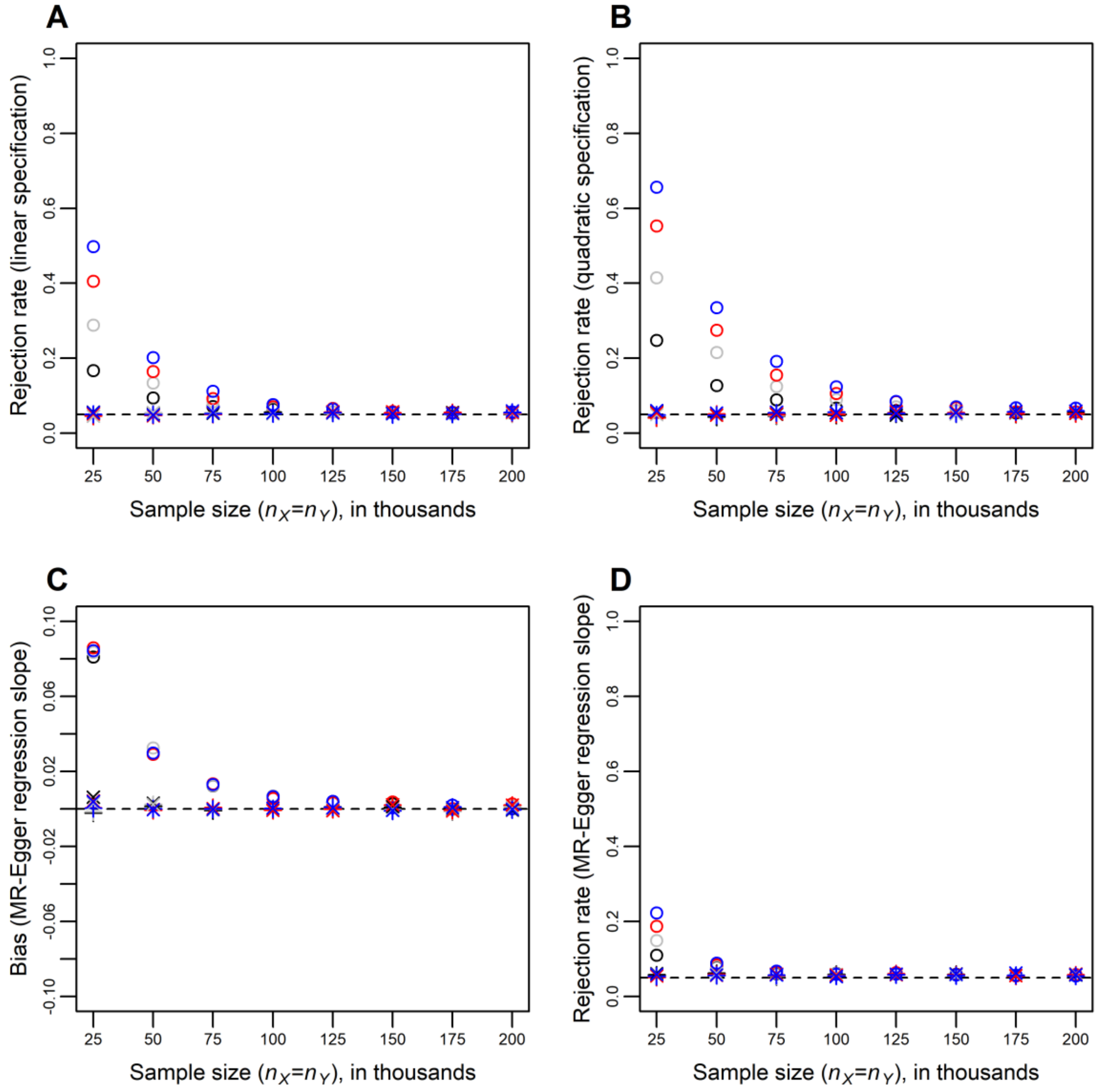